\documentclass[aps,reprint,twocolumns,pre,superscriptaddress,floatfix,notitlepage,nofootinbib]{revtex4-1}
\usepackage{amssymb,amsmath,amsfonts} 
\usepackage{mathtools}
\usepackage{physics}
\usepackage{array}
\usepackage{graphicx,epsfig,xcolor}
\usepackage[colorlinks=true,citecolor=blue,linkcolor=red]{hyperref}
\usepackage{newtxtext,newtxmath}

\newcommand{\hatmath}[1]{\hat{\mathcal{#1}}} 
\newcommand{\beq}{\begin{equation}}
\newcommand{\eeq}{\end{equation}}

\begin{document}

\title{Dynamical and excited-state quantum phase transitions in collective systems}

\author{\'{A}ngel L. Corps}
    \email[]{corps.angel.l@gmail.com}
    \affiliation{Instituto de Estructura de la Materia, IEM-CSIC, Serrano 123, E-28006 Madrid, Spain}
    \affiliation{Grupo Interdisciplinar de Sistemas Complejos (GISC),
Universidad Complutense de Madrid, Av. Complutense s/n, E-28040 Madrid, Spain}
    
\author{Armando Rela\~{n}o}
    \email[]{armando.relano@fis.ucm.es}
    \affiliation{Grupo Interdisciplinar de Sistemas Complejos (GISC),
Universidad Complutense de Madrid, Av. Complutense s/n, E-28040 Madrid, Spain}
    \affiliation{Departamento de Estructura de la Materia, F\'{i}sica T\'{e}rmica y Electr\'{o}nica, Universidad Complutense de Madrid, Av. Complutense s/n, E-28040 Madrid, Spain}

\date{\today} 

\begin{abstract}
We study dynamical phase transitions (DPTs) in quantum many-body systems with infinite-range interaction, and present a theory connecting the two kinds of known DPTs (sometimes referred to as DPTs-I and DPTs-II) with the concept of excited-state quantum phase transition (ESQPT), traditionally found in collective models. We show that DPTs-I appear as a manifestation of symmetry restoration after a quench from the broken-symmetry phase, the limits between these two phases being demarcated precisely by an ESQPT. We describe the order parameters of DPTs-I with a generalization of the standard microcanonical ensemble incorporating the information of two additional conserved charges identifying the corresponding phase. We also show that DPTs-I are linked to a mechanism of information erasure brought about by the ESQPT, and quantify this information loss with the statistical ensemble that we propose. Finally, we show analytically the main mechanism for DPTs-II is forbidden in these systems for quenches leading a broken-symmetry initial state to the same broken-symmetry phase, on one side of the ESQPT, and we provide a formulation of DPTs-II depending on the side of the ESQPT where the quench ends. We analyze the connections between various indicators of DPTs-II. Our results are numerically illustrated in the infinite-range transverse-field Ising model and are applicable to a large class of collective quantum systems satisfying a set of conditions.
\end{abstract}

\maketitle

\section{Introduction}

Unveiling new exotic phases of quantum many-body systems is one of the primary goals of modern research in quantum physics. This endeavor has been notably encouraged by state-of-the-art techniques involving cold atoms and trapped
ions \cite{Gring2012, Hofferberth2007, Hild2014, Yuznashyan2006,Baumann2010,Muniz2020,Chu2020}, which are used both to precisely simulate those systems in a laboratory and also in the search for new physics. 

During recent years the term dynamical quantum phase transition (DPT) has been used to denote two different phenomena. The first type, DPTs-I, is characterized by a non-equilibrium order parameter which changes non-analytically at a critical point that separates two dynamical phases \cite{Eckstein2008,Moeckel2008,Eckstein2009,Sciolla2011,Zhang2017,Muniz2020,Smale2019,Tian2020,Halimeh2017prethermalization,Sciolla2013,Sciolla2011,Alvarez2006}. It is usually triggered by a quantum quench ---a sudden change in a control parameter--- which takes the system out of equilibrium. It is normally linked to pre-thermalization, that is, to a long-lived steady state in which the system remains during long times \cite{Gring2012,Mori2018}, and, in many cases, it gives rise to persistent oscillations.

Similar dynamical phases appear in classical open systems, which remain out of equilibrium due to the interaction with their environments; typical examples are oscillating chemical reactions, convection cells or cyclic patterns in population dynamics \cite{Prigogine1971}. In the quantum domain, the same phenomenon appears in closed quantum systems evolving under unitary dynamics. A typical scenario occurs in quantum systems with long-range or infinite-range interactions which undergo a quantum phase transition (QPT) separating two ground-state phases \cite{Marino2022}: one phase where a discrete $\mathbb{Z}_2$ symmetry is broken, and another where the same symmetry is restored. Under such circumstances, a quench from a broken-symmetry ground state may lead to two different dynamical behaviors: one in which oscillations around a broken-symmetry effective state are observed, and another one in which the same kind of oscillations occur around a symmetric state. In the first case, the dynamical order parameter may be different from zero; in the second, it is always equal to zero. Both phases are separated by a critical quench leading the system to a particular value of the control parameter which, in general, does \textit{not} coincide with the critical coupling of the QPT  \cite{Sciolla2011}. At the critical quench, the dynamical order parameter changes non-analytically in the thermodynamic limit (TL).

The second phenomenon, DPTs-II \cite{Heyl2018}, also happens in closed quantum systems evolving under unitary dynamics. However, it is not characterized by a dynamical order parameter, but by the appearance of critical times when the return probability becomes non-analytic
\cite{Heyl2013,Heyl2014,Jurcevic2017,Homrighausen2017,Halimeh2017,Heyl2019,Nicola2021,Schmitt2015,Bhattacharya2017,Karrasch2013,Peng2015,Jafari2022,Naji2022,Jafari2019,Mishra2020,Jafari2019prb,Jafari2021}. It was originally identified in a one-dimensional transverse-field Ising chain with nearest-neighbor interactions \cite{Heyl2013}, taking advantage of the mathematical equivalence between a boundary partition function and the return probability of a time-evolving wavefunction. The resulting critical time is signaled by a non-analytical point in the return probability in the TL, much in the same way that a critical temperature is identified by a non-analytical point in the partition function, also in the TL. In general, DPTs-II are unrelated to equilibrium phase transitions \cite{Vajna2014,Andraschko2014,Jafari2019}, and they seem to depend strongly on the initial condition chosen \cite{Lang2018concurrence,Homrighausen2017,Halimeh2017}. This kind of DPT also occurs in quantum systems with long-range or infinite-range interactions exhibiting the same kinds of QPTs that are linked to DPTs-I. Indeed, it is known that both DPTs-I and DPTs-II can take place in the same models, like the fully connected transverse-field Ising model \cite{Zunkovic2018,Lang2018}, or the Rabi model \cite{Puebla2020}. Connections between both kinds of DPTs have been proposed \cite{Zunkovic2018,Lang2018,Weidinger2017,Hashizume2022,Sehrawat2021,Zunkovic2015,Lerose2019,Lang2018concurrence}; nevertheless, a common triggering mechanism has not been found.

In this paper we discuss a theory that links these two kinds of DPTs to excited-state quantum phase transitions (ESQPT) \cite{Cejnar2021} in collective quantum systems, and illustrate it by means of numerical results on a fully-connected transverse-field Ising model. This theory is proposed in \cite{Corps2022letter}, for which this work serves as a companion paper. There, it is shown that both kinds of DPTs are explained by the behavior of an operator, $\hat{{\mathcal C}}$, that is a constant of motion only in one of the two phases separated by the critical energy of the ESQPT \cite{Corps2021}. Below this critical energy, $\hat{{\mathcal C}}$ commutes with the energy projectors in the TL, the dynamical order parameter characteristic of DPTs-I can be different from zero, and the main mechanism leading to non-analytical points in the return probability is precluded. Contrarily, above the ESQPT critical energy, the same dynamical order parameter is always equal to zero, and the main mechanism for non-analytic points in the return probability is allowed. We discuss the semiclassical basis of this theory, and study some of its consequences, like the suppression of the semiclassical oscillations for critical quenches, the information erasure due to the adiabatic crossing of the critical energy of the ESQPT, and the differences between the anomalous and the normal dynamical phases linked to DPTs-II \cite{Homrighausen2017,Halimeh2020}. To make the paper self-contained, we also review the main results presented in \cite{Corps2022letter}. 

This paper is organized as follows. In Sec. \ref{sec:setup} we set the scene by detailing the general properties of the collective systems to which our results apply. In Sec. \ref{sec:model} we review the transverse-field Ising model with infinite-range interactions, which we use to illustrate our results. Sec. \ref{sec:semiclassical} is devoted to a semiclassical analysis of this system, which proves very fruitful to understand some of our results. In Sec. \ref{sec:dpti} we focus on DPTs-I. In particular, in Sec. \ref{sec:quantumquenches} and Sec. \ref{sec:timeevolution} we study the dynamics of order parameters of DPTs-I after a quench, and we develop a statistical ensemble to describe the long-time averages around which such time-evolved expectation values oscillate in Sec. \ref{sec:gme}. Then, in Sec. \ref{sec:timedependent} we study the adiabatic dynamics of these order parameters. We focus on DPTs-II in Sec. \ref{sec:dptii}. We show analytically that the main mechanism for DPTs-II is not allowed in one of the phases demarcated by the ESQPT in Sec. \ref{sec:analyticalargument}, and that the sum of the return probabilities to each of the parity-broken ground states coincides with the survival probability in the same region of the ESQPT. These DPTs-II are further explored numerically in Sec. \ref{sec:numericalresults}, and we discuss the results in Sec. \ref{sec:discussiondptii}, where we propose a formulation of DPTs-II in terms of the energy of the quenched state (above or below the ESQPT), commenting on some open problems. Finally, we conclude in Sec. \ref{sec:conclusions}.

\section{Generic setup}\label{sec:setup}

Although all numerical results in this paper concern the fully-connected transverse-field Ising model, our theory is applicable to a broad class of collective quantum systems; as paradigmatic examples, we highlight the Lipkin-Meshkov-Glick model (LMG) model (a version of which is mathematically equivalent to the fully-connected transverse-field Ising model) \cite{Lipkin1965,Dusuel2004,Heiss2005,Leyvraz2005,Castanos2006,Ribeiro2007,Ribeiro2008,Relano2008,GarciaRamos2017}, the Rabi and Dicke models \cite{Relano2016,Puebla2016,Hwang2015,Lobez2016,Bastarrachea2014,Perez2011b,Perez2011,Brandes2013,LewisSwan2021,Kloc2017,Kloc2018}, the coupled top \cite{Wang2021}, spinor Bose-Einstein condensates \cite{Feldmann2021}, or the two-site Bose-Hubbard Hamiltonian \cite{Relano2014}, to cite a few.

Let us consider a Hamiltonian $\hatmath{H}(\lambda)$, depending on some
control parameter, $\lambda$, with the following properties:

(i) There exists a discrete ${\mathbb Z}_2$
symmetry, represented by a discrete operator, $\hat{\Pi}$, which we call parity,
fulfilling $[\hatmath{H}(\lambda),\hat{\Pi}]=0$, $\forall
\lambda$, allowing to classify the eigenstates of $\hatmath{H}(\lambda)$, $\ket{E_{n,k}}$, according to $\hat{\Pi}\ket{E_{n,k}}=k\ket{E_{n,k}}$ with $k=\pm1$ and $n=0,1,2,...$. Typical examples are the inversion of the transverse magnetic field in the fully-connected Ising model, $\hat{J}_x \rightarrow - \hat{J}_x$, or the same mathematical transformation together with a similar one for the bosonic part of the Hilbert space, $\hat{a} \rightarrow -\hat{a}$, in the Dicke and Rabi models. 

(ii) There exists a critical value for the control parameter, $\lambda_c$, at which a QPT \cite{Sachdev1999} occurs. This critical point separates two ground-state phases. On one side, say $\lambda > \lambda_c$, the ground-state is two-fold degenerate in the TL (pairs of eigenvalues with different parity coincide, $E_{n,+}=E_{n,-}$), and therefore it becomes possible to find a broken-symmetry ground-state. It corresponds to the ferromagnetic ordered phase in the fully connected transverse-field Ising model, or to the superradiant phase of the Dicke and Rabi models. On the other side, say $\lambda<\lambda_{c}$, the ground state is unique, and it has a well-defined value of the parity symmetry. It corresponds to the disordered paramagnetic phase in the fully-connected transverse-field Ising model, or to the normal phase in the Dicke and Rabi models. This is the typical scenario for DPTs-I, as we have pointed out in the Introduction.

(iii) In the ordered phase, $\lambda > \lambda_c$, some properties of the ground state are extended up to a certain excited critical energy, e.g. $E_{\textrm{gs}}<E<E_c$, at which a ESQPT takes place. For example, all the energy levels become two-fold degenerate below this critical energy, $E_{c}$, in the TL \cite{Cejnar2021,Corps2021}, and therefore broken-symmetry equilibrium or steady states are allowed in this region \cite{Puebla2013,Puebla2013b,Puebla2015}. Contrarily, for $E>E_c$ all degeneracies are broken and thus broken-symmetry equilibrium states are no longer possible. These phases are separated by a singularity in the density of states at $E=E_{c}$ and in the energy level flow. The character of this singularity ultimately depends on the properties of the semiclassical limit of the quantum system. If the system has a single semiclassical degree of freedom, then the density of states, $\varrho(E)$, may commonly show a logarithmic singularity at $E_c$. If it has two, the logarithmic singularity is transferred onto the first derivative of the density of states, and so on \cite{Cejnar2021,Stransky2014}. As the number of classical degrees of freedom increases, the signatures of the ESQPT may be harder to find. 

\begin{center}
\begin{figure}[h!]
\hspace*{-0cm}\includegraphics[width=0.52\textwidth]{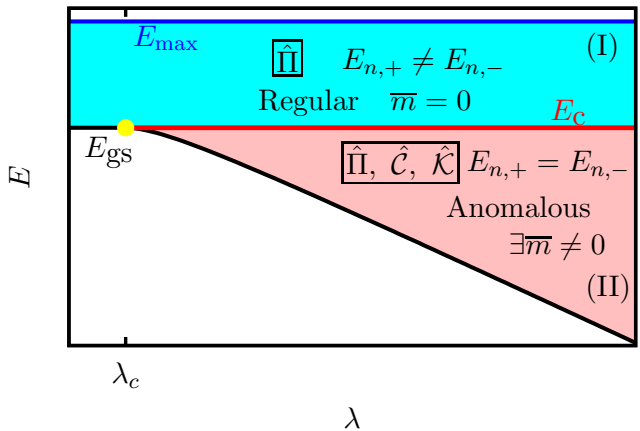}
\caption{Schematic representation of the phase diagram of the class of systems to which the results of this paper apply. Energy of the model eigenstates is represented versus the control parameter $\lambda$. The black lower line represents the ground state of the model, $E_{\textrm{gs}}$. A QPT occurs at $\lambda=\lambda_{c}$ (yellow circle). When $\lambda>\lambda_{c}$, an ESQPT takes place at energy $E_{c}$ (thick red line). The maximum energy, where the phase diagram ends, is $E_{\max}$ (upper blue line). Two phases emerge (I and II) characterized by different symmetries (squared, see main text). In (I) the $n$th states of different parity, $E_{n,\pm}$, have different energy, $E_{n,+}\neq E_{n,-}$, while in (II) they are degenerate, $E_{n,+}=E_{n,-}$ and the parity symmetry is broken. Regarding DPTs-I,the phase (I) is characterized by a vanishing order parameter, $\overline{m}=0$, while in (II) this order parameter can take a non-zero value, $\overline{m}\neq 0$. Regarding DPTs-II, when starting from an initial, broken-symmetry state from (II) a quench to (I) will give rise to a regular phase while a quench to (II) will reveal an anomalous phase (see main text and discussions). }
\label{fig:phasediagram}
\end{figure}
\end{center}

For the purposes of this paper, the trademark of the phase where $\lambda > \lambda_c$ and $E<E_c$ is the existence of an operator $\hat{{\mathcal C}}$ that becomes a constant of motion in the TL, proposed in \cite{Corps2021}. We have that $\left[ \hat{{\mathcal C}}, \hat{P}_n \right]=0$, where $\hat{P}_n$ is the projector onto the eigenspace with energy $E_n$, $\forall E_n < E_c$, whereas $\left[ \hat{{\mathcal C}}, \hat{P}_n \right] \neq 0$, $\forall E_n > E_c$. Just like $\hat{\Pi}$, this operator has only two different eigenvalues, $\textrm{Spec} \,(\hat{{\mathcal C}})=\{\pm 1\}$, and acts like a partial $\mathbb{Z}_2$ symmetry. Furthermore, it does {\em not} commute with the parity symmetry, $[\hatmath{C},\hat{\Pi}]\neq0$. As we will discuss in Sec. \ref{sec:dpti}, this immediately implies the existence of a third constant of motion in this phase, $\hatmath{K}$. 

A schematic representation of these properties is in Fig. \ref{fig:phasediagram}. We represent a phase diagram for DPTs-I and DPTs-II in the form of a level-flow diagram (i.e., the energy of the model eigenstates $E$ as a function of the control parameter $\lambda$). Two distinct dynamical phases emerge, (I) and (II). Regarding DPTs-I, starting from an initial broken-symmetry state in (II) and quenching it to (I), there exists dynamical order parameters which always vanish, $\overline{m}=0$; this is a consequence of the operators $\hatmath{C}$ and $\hatmath{K}$ not being constant in this phase. However, a quench leading the initial state from (II) to (II) also will produce dynamical order parameters that may be non-zero, $\overline{m}\neq 0$ as a consequence of the conservation of $\hatmath{C}$ and $\hatmath{K}$ (for specific initial states it may also be that $\overline{m}=0$, but this is not the general case in this phase).  Regarding DPTs-II, the same quench protocol will reveal a regular phase in (I) and an anomalous phase in (II) (for more details see below). These two phases are separated by the ESQPT critical energy, $E_{c}$, which may or may not be a function of $\lambda$. The maximum energy of the system, $E_{\max}$, may be finite in models such as the LMG, $E_{\max}<\infty$, or it may be infinite in other systems where there is a coupling to photonic degrees of freedom, like the Rabi or Dicke models, $E_{\max}\to\infty$.

\section{Model}\label{sec:model}

As mentioned, to illustrate our results we use the infinite-range transverse-field Ising model \cite{Sachdev1999}, which coincides with a simple version of the well-known LMG Hamiltonian from nuclear physics \cite{Lipkin1965},
\begin{equation}\label{eq:lipkin}
\hatmath{H}(\lambda)=-\frac{\lambda}{4N}\sum_{i,j=1}^N \hat{\sigma}^x_i \hat{\sigma}^x_j + \frac{h}{2} \sum_{i=1}^N \hat{\sigma}^z_i= -\frac{\lambda}{N}\hat{J}_{x}^{2}+h \hat{J}_{z},
\end{equation}
where $\hbar =1$. Here, $\hat{\sigma}_{x,y,z}$ are the Pauli matrices, and $\hat{\mathbf{J}}=(\hat{J}_{x},\hat{J}_{y},\hat{J}_{z})$ are the collective spin operators for the $N$ $1/2$-spins that form the system, $\hat{J}_{k}=\frac{1}{2}\sum_{i=1}^{N}\hat{\sigma}_{i}^{k}$ ($k\in\{x,y,z\}$). The total spin operator $\mathbf{\hat{J}}^{2}$ is an exact conserved quantity, $[\hatmath{H}(\lambda),\hat{\mathbf{J}}^{2}]=0$. Hence we can separate the Hamiltonian matrix in symmetry sectors according to its eigenvalues, $j(j+1)$; we focus on the maximally symmetric sector, $j=N/2$. The collective spin length $j$ is the parameter leading to the TL (see below). The parity $\hat{\Pi}\equiv e^{i\pi(j+\hat{J}_{z})}$ is also an exact discrete conserved quantity with two eigenvalues, $\textrm{Spec}\,(\hat{\Pi})=\{\pm 1\}$, so the eigenstates of Eq. \eqref{eq:lipkin} can be classified according to $\hat{\Pi}\ket{E_{n,\pm}}=\pm \ket{E_{n,\pm}}$.  For our numerical calculations we will fix $h=1$ and consider $\lambda$ as the single control parameter. In recent experimental realizations \cite{Zhang2017,Muniz2020,Jurcevic2017} $\lambda$ is kept fixed and $h$ is allowed to vary. Our choice is equivalent to this procedure, and it allows a more clear identification of the critical points of the system (see below). Also, in \cite{Muniz2020} the magnetic field is on the scale $h\sim$ MHz; we adopt this scale, so in our numerical results, $t\sim $ $\mu$s. 

The structure of QPTs and ESQPTs of Eq. \eqref{eq:lipkin} is analyzed for completeness in Sec. \ref{sec:semiclassical}. The model has a QPT at $\lambda_{c}=h$ \cite{Botet1982,Ribeiro2007,Ribeiro2008}. For $\lambda>\lambda_{c}$, it also exhibits an ESQPT exactly at $ E_{c}=-hj$, or $\epsilon_{c}\equiv E_{c}/j=-h$, commonly signaled by a logarithmic singularity in the level density \cite{Dusuel2004,Heiss2005,Leyvraz2005,Castanos2006,Ribeiro2007,Ribeiro2008,Relano2008}. For $\lambda>\lambda_{c}$ and $E<E_{c}$, the LMG model displays a broken-symmetry phase where the eigenstates of different parity are degenerate, $E_{n,+}=E_{n,-}$. However, if either $\lambda<\lambda_{c}$, or $\lambda>\lambda_{c}$ and $E>E_{c}$, this degeneracy is lifted, $E_{n,+}\neq E_{n,+}$ and the $\mathbb{Z}_{2}$ symmetry is restored. Thus it is clear that this model belongs to the family of systems described in Sec. \ref{sec:setup}.

\subsection{Semiclassical analysis}\label{sec:semiclassical}
The Hamiltonian Eq. \eqref{eq:lipkin} represents a \textit{collective} system, where each spin interacts with every other. The number of degrees of freedom remains finite when the collective spin length increases boundlessly, and thus the \textit{thermodynamic} limit $j\to\infty$ coincides with a \textit{semiclassical} limit, $\hbar\to 0$ \cite{Cejnar2021}. This mean-field solution can be obtained, e.g., by using the Bloch coherent state associated with the SU(2) group,

\beq\label{eq:coherent} \ket{\omega}=\left(\frac{1}{1+|\omega|^{2}}\right)^{j}e^{\omega\hat{J}_{+}}\ket{j,-j},
\eeq
where $\ket{j,-j}$ is the state with spin $j$ and $\langle \hat{J}_{z}\rangle =-j$, and
\beq
\omega=\frac{Q+iP}{\sqrt{4-P^{2}-Q^{2}}}\in\mathbb{C}
\eeq 
with $Q$ and $P$ real variables. Then, we take the expectation value of the quantum Hamiltonian $\hatmath{H}$ in Eq. \eqref{eq:coherent}. On the scale of the collective spin length $j$, this gives the intensive energy functional 

\beq\label{eq:classicalham}
H(Q,P;\lambda)\equiv \frac{\bra{\omega}\hat{\mathcal{H}}\ket{\omega}}{j}=-h+\frac{h}{2}(Q^{2}+P^{2})-\frac{\lambda}{8}Q^{2}(4-P^{2}-Q^{2}).
\eeq
Here, $(Q,P)$ are canonical variables constrained to a 2-dimensional ball of radius 2, i.e., the phase space is $\mathcal{M}=\{(Q,P)\in\mathbb{R}^{2}\,\,:\,\,0\leq Q^{2}+P^{2}\leq 4\}$. Clearly, the classical model Eq. \eqref{eq:classicalham} has a single degree of freedom, $f=1$. To allow a convenient comparison with the quantum Hamiltonian, we define the intensive energy scale associated with the classical Hamiltonian $\epsilon\equiv E/j$, where $E$ denotes the actual eigenvalues of the Hamiltonian Eq. \eqref{eq:lipkin} and $\epsilon$ refers to the scale in Eq. \eqref{eq:classicalham}. $E$ and $\epsilon$ will be used depending on which of these two energy scales we are referring to. 

The Bloch coherent states make it possible to obtain a classical representation of common dynamical functions too. For example, the classical variable for the collective magnetization $j_{z}=\bra{\omega}\hat{J}_{z}\ket{\omega}/j$ is 
\begin{equation}
j_{z}=\frac{Q^{2}+P^{2}}{2}-1,
\end{equation}
while the parity-breaking spin operator $j_{x}=\bra{\omega}\hat{J}_{x}\ket{\omega}/j$ reads 

\beq\label{eq:jxclassical}
j_{x}=\frac{Q}{2}\sqrt{4-P^{2}-Q^{2}}.
\eeq
Knowledge about the structure and phase transitions present in the classical model can be gained by analyzing the \textit{fixed points} of Eq. \eqref{eq:classicalham} \cite{Cejnar2021},  $\eval{\nabla H}_{(Q*,P*)}=0$, which in turn coincide with \textit{stationary points} of the dynamics,

\beq\label{eq:partialP}
\frac{\partial H}{\partial P}=P\left(h+\frac{\lambda}{4}Q^{2}\right)=\frac{\textrm{d}Q}{\textrm{d}t},
\eeq
\beq\label{eq:partialQ}
\frac{\partial H}{\partial Q}=-
Q\left[\left(\lambda-h\right)+\frac{\lambda}{4}\left(-P^{2}-2Q^{2}\right)\right]=-\frac{\textrm{d}P}{\textrm{d}t}.
\eeq
Nullification of Eqs. (\ref{eq:partialP},\ref{eq:partialQ}) provides different real solutions depending on the value of $\lambda\geq0$, and all solutions are of the form $(Q,P)=(Q,0)$. If $\lambda<h$, the only critical point has $Q=0$, corresponding to $\epsilon=-h$. This is the ground-state energy when $\lambda<h$. However, if $\lambda\geq h$, there exist two additional critical points, $Q=\pm \sqrt{2(\lambda-h)/\lambda}$. A second order QPT occurs at the critical value of the control parameter $\lambda_{c}=h$.  Since the classical Hamiltonian Eq. \eqref{eq:classicalham} exhibits the symmetry $H(Q,0)=H(-Q,0)$, these two critical points correspond to the exact same energy, $\epsilon=-(h^{2}+\lambda^{2})/2\lambda$, which is the ground-state energy if $\lambda\geq h$. The previous critical point with $Q=0$ corresponds to an unstable fixed point if $\lambda\geq h$, defining an ESQPT at $\epsilon_{c}=-h$, $\forall \lambda>\lambda_{c}$. After this general discussion, we will fix $h=1$ in our numerical results and consider $\lambda$ as the single control parameter.

\begin{center}
\begin{figure}[h!]
\hspace*{-0.6cm}\includegraphics[width=0.52\textwidth]{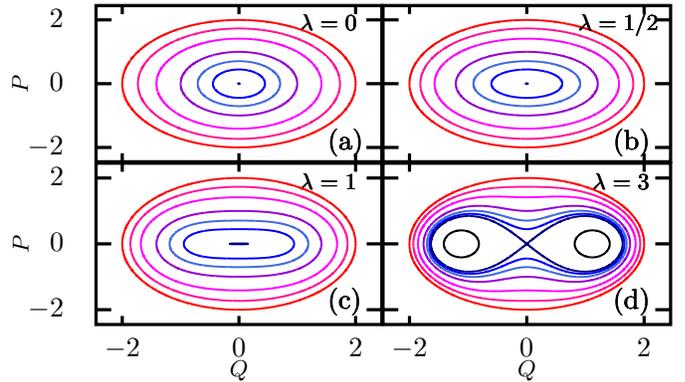}
\caption{Classical phase space of the Hamiltonian Eq. \eqref{eq:classicalham}. In (a-d) the control parameter $\lambda$ is varied, and $h=1$. Lines represent different classical orbits at different energies; each color is linked to a given energy. Magenta-red lines show higher energies, while purple-blue lines show lower energies.}
\label{fig:phasespace}
\end{figure}
\end{center}

Fig. \ref{fig:phasespace} illustrates the structure of the classical phase space by means of classical orbits. Each line represents the set of points $(Q,P)$ such that $H(Q,P)=\epsilon$. For $\lambda=0$ the LMG model reduces the trivial Hamiltonian $\hatmath{H}=-h\hat{J}_{z}$ and thus the phase space is essentially that of a harmonic oscillator, $H=-h+h(Q^{2}+P^{2})/2$, composed of concentric circumferences. We note that the ground-state is unique and no relevant feature can be observed. A value $\lambda\neq 0$ introduces distortions with respect to the perfect harmonic behavior, but the ground-state is still unique as shown in Fig. \ref{fig:phasespace}(b-c). A representative picture of the phase space when $\lambda>\lambda_{c}$ is shown in Fig. \ref{fig:phasespace}(d) for $\lambda=3$. There are three relevant features: (i) the ground-state is pairwise degenerate, occurring at a given $Q$ and also at its mirrored image, $-Q$; (ii) there appears an unstable fixed point at $(0,0)$ (where the orbit appears to `cross itself') at energy $\epsilon_{c}=-1$ (for $h=1$); and (iii) for $\epsilon<\epsilon_{c}$ trajectories are trapped within either the right or left classical wells, depending on the initial condition, but if $\epsilon>\epsilon_{c}$ there is no such constraint and trajectories can explore all available phase space.

\begin{center}
\begin{figure}[h!]
\hspace*{-0.52cm}\includegraphics[width=0.5\textwidth]{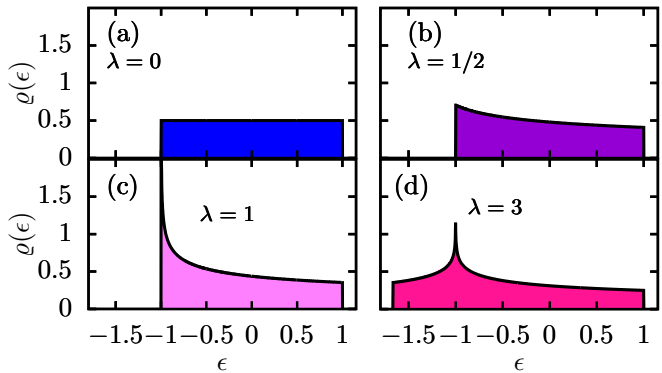}
\caption{Density of states, $\varrho(\epsilon)$, for different values of the control parameter $\lambda$, and $h=1$. Color filled histograms represent the numerical level density obtained from the quantum model, Eq. \eqref{eq:lipkin}, with system size $j=10000$, while black lines show the semiclassical approximation to the level density, Eq. \eqref{eq:classicaldensity}.}
\label{fig:paneldensity}
\end{figure}
\end{center}

The energy $\epsilon_{c}=-1$ at which trajectories display this `singular' behavior is associated to an ESQPT, evidenced by the classical level density \cite{Stransky2014,Caprio2008}. This can be approximated by the first term of Gutzwiller's trace formula \cite{Gutzwillerbook}, namely ($\hbar =1$)

\beq\label{eq:classicaldensity}
\varrho(\epsilon)=\frac{1}{2\pi}\int_{\mathcal{M}} \textrm{d}Q\textrm{d}P\,\delta [\epsilon-H(Q,P)]
\eeq
which is displayed in Fig. \ref{fig:paneldensity} for the same values of $\lambda$ as in Fig. \ref{fig:phasespace}. When $\lambda=0$ the spectrum is equiespaced and, thus, $\varrho(\epsilon)$ is uniform. A non-zero $\lambda$ distorts this shape and brings about a ramp-shaped distribution with a small peak at the border, $\epsilon=-1$. This is transformed into a \textit{logarithmic singularity} when $\lambda\geq\lambda_{c}$, as shown for $\lambda=1$ and $\lambda=3$. 

It is clear from Fig. \ref{fig:phasespace} that for control parameter $\lambda>\lambda_{c}$ and for energies $\epsilon\leq \epsilon_{c}$, the classical dynamics conserves the sign of the canonical variable $Q(t)$, as trajectories are trapped within one of the two accessible energy wells [cf. Fig. \ref{fig:phasespace}(d)]. This stems from the disconnectedness of the phase space topology in this case. However, this conservation is no longer present when either $\lambda<\lambda_{c}$ or $\lambda>\lambda_{c}$ and $\epsilon>\epsilon_{c}$, because then the phase space is topologically connected and a given trajectory has access to all available phase space [cf. Fig. \ref{fig:phasespace}(a-c)]. It has been recently proposed \cite{Corps2021,Corps2022,Corps2022arxiv} the classical constant of motion $\textrm{sign}\,[Q(t)]$ can be translated to the quantum domain by establishing a connection between quantum and classical dynamical functions. In this case, $(Q,P)$ are bounded to the circumference of radius 2, $\sqrt{4-P^{2}-Q^{2}}\geq 0$, and therefore Eq. \eqref{eq:jxclassical} implies that below the ESQPT critical energy, $j_{x}$ can only be positive or negative depending on whether the initial conditions lie in the right or left classical wells, respectively, i.e., $\textrm{sign}\,(Q)=\textrm{sign} (j_{x})$. Thus, the quantum operator providing conserved quantum numbers when $\lambda>\lambda_{c}$ and $\epsilon<\epsilon_{c}$ is simply

\beq\label{eq:constant}
\hatmath{C}=\textrm{sign}\,(\hat{J}_{x}).
\eeq
Eq. \eqref{eq:constant} defines a discrete $\mathbb{Z}_{2}$ symmetry with only two eigenvalues, $\textrm{Spec}\,(\hatmath{C})=\{\pm 1\}$ \cite{Corps2021}, and it is an instance of a so-called partial symmetry because it is only commuting with the projectors onto the Hamiltonian eigenspaces corresponding to energy below the ESQPT criticality, $\epsilon<\epsilon_{c}$. As shown in \cite{Corps2021}, for eigenstates $\ket{E_{n,\pm}}$ with energy below the ESQPT critical energy, the eigenvectors of $\hatmath{C}$ are $(\ket{E_{n,+}}\pm \ket{E_{n,-}})/\sqrt{2}$, and its diagonal expectation values are $\bra{E_{n,\pm}}\hatmath{C}\ket{E_{n,\mp}}=\pm 1$, where the overall sign in the right-hand side is not related to parity but to the arbitrary global phases of each of the eigenstates $\ket{E_{n,\pm}}$, and therefore it may be fixed to $+1$ for definiteness \cite{Corps2021}. This operator establishes a connection between quantum states and the region of phase space where they should be classically attached to; specifically, the expectation value of $\hatmath{C}$ in an arbitrary state $\ket{\varphi(t)}$, $\langle\hatmath{C}(t)\rangle=\bra{\varphi(t)}\hatmath{C}\ket{\varphi(t)}\in[-1,1]$, indicates whether the quantum state is attached to the left (-1) or right (+1) classical energy wells, or if a superposition of these two limits occurs (between -1 and +1). If the state $\ket{\varphi(t)}$ does not populate Hamiltonian eigenstates with energy $E>E_{c}$, then such a expectation value remains constant in time. We will  use these ideas in the following sections.

\section{DPT-I: Dynamical order parameter}\label{sec:dpti}

As advanced in the Introduction, DPTs-I are non-analyticities characterized by a non-equilibrium order parameter after a quantum quench \cite{Eckstein2008,Moeckel2008,Eckstein2009,Sciolla2011,Zhang2017,Muniz2020,Smale2019,Tian2020,Halimeh2017prethermalization,Sciolla2013,Sciolla2011}. In this section we will reveal how DPTs-I are closely connected with ESQPTs in a large class of \textit{collective} many-body quantum systems: we will show how DPTs-I stem from a symmetry restoration brought about by the ESQPT and will quantify the long-time average of order parameters with a generalization of the standard microcanonical ensemble \cite{Alessio2016}. As we will see, this statistical ensemble contains the information of three noncommuting charges directly related to the ESQPT non-analyticity. 

\subsection{Quantum quenches}\label{sec:quantumquenches}

\begin{center}
\begin{figure}[h]
\hspace*{-0.65cm}\includegraphics[width=0.52\textwidth]{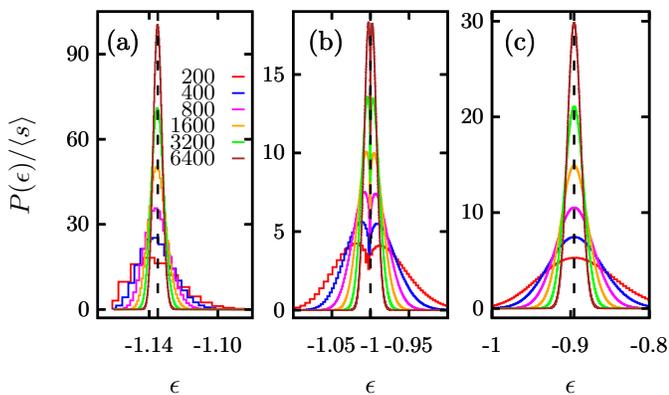}
\caption{(a-c) Probability of populated states after a quench $\lambda_{i}\to\lambda_{f}=1.75$ from the initial state Eq. \eqref{eq:initialstate} with $\alpha=1/2$ and $\phi=0$. The values of the initial control parameter and the quenched state energy are (a) $\lambda_{i}=2.5$, $\epsilon(\lambda_{f})=-1.135$ (b) $\lambda_{i}=7$ , $\epsilon(\lambda_{f})=-1$, and (c) $\lambda_{i}=27.5$, $\epsilon(\lambda_{f})=-0.89567$. The system size $j$ ranges from 200 to 6400 as indicated in (a). The black dashed lines represent the average energy of the quench in the TL. }
\label{fig:ldosmeanspacing}
\end{figure}
\end{center}

As an initial state, we start from a superposition of the broken-symmetry ground state at an initial value of the control parameter, $\lambda_{i}>\lambda_{c}$,  
\begin{equation}\label{eq:initialstate}
    \ket{\Psi_{0}(\lambda_{i})}=\sqrt{\alpha}\ket{E_{0,+}(\lambda_{i})}+e^{i\phi}\sqrt{1-\alpha}\ket{E_{0,-}(\lambda_{i})},
\end{equation}
where $\alpha\in[0,1]$, $\phi\in[0,2\pi)$, and $\ket{E_{0,\pm}(\lambda_{i})}$ denotes the ground-state eigenstate of the initial Hamiltonian, $\hatmath{H}(\lambda_{i})$, with parity $\pm 1$. Recent experimental realizations  \cite{Zhang2017,Muniz2020,Jurcevic2017} follow the same protocol with $\alpha=1/2$ and $\phi=0$. Then, we quench the system to $\lambda_{f}$, and allow a unitary time evolution in the final Hamiltonian ($\hbar=1$),

\beq\begin{split}
&\ket{\Psi_{t}(\lambda_{f})}=e^{-i\hatmath{H}(\lambda_{f})t}\ket{\Psi_{0}(\lambda_{i})}\\&=\sum_{n}\sum_{k=\pm}\bra{E_{n,k}(\lambda_{f})}\ket{\Psi_{0}(\lambda_{i})}e^{-i E_{n,k}(\lambda_{f})t}\ket{E_{n,k}(\lambda_{f})}.
\end{split}
\eeq
The distribution of populated states after the quench is

\begin{equation}\label{eq:populationquench}
P(E)=\sum_{n}\sum_{k=\pm}|c_{n,k}|^{2}\delta(E-E_{n,k}),
\end{equation}
where the coefficients $c_{n,k}\equiv \bra{E_{n,k}(\lambda_{f})}\ket{\Psi_{0}(\lambda_{i})}$. This distribution is represented in Fig. \ref{fig:ldosmeanspacing} for several values of the collective spin-length, $j$, approaching the TL, starting from different values of $\lambda_{i}>\lambda_{c}$, $\alpha=1/2$ and $\phi=0$, and all finishing at $\lambda_{f}=1.75$. Depending on $\lambda_{i}$, the average energy of the quenched state, $\sum_{n}\sum_{k=\pm}|c_{n,k}|^{2}\epsilon_{n,k}(\lambda_{f})$, may be driven from one side of the ESQPT, $E<E_{c}$, to the other, $E>E_{c}$. The distribution is scaled by the mean level spacing, $\langle s\rangle$ where the level spacing, $s_{n}=\epsilon_{n+1}-\epsilon_{n}$, shows a typical decrease as $j$ increases (the rescaled energy spectrum $\{\epsilon_n\}$ becomes denser as $j$ increases).  The width of the distribution shrinks with $j$ and the distribution becomes more peaked precisely around the infinite-$j$ average given by the semiclassical model, indicated in all panels with a vertical dashed line. Importantly, the distributions in (a) and (c) are well located within a definite side of the ESQPT: in (a) $P(\epsilon)$ shows significant population only below $\epsilon_{c}=-1$, while in (c) the opposite happens. However, in (b) the average energy of the quenched state coincides with $\epsilon_{c}=-1$. One can see that both sides of the ESQPT are significantly populated, with a clear dip of the distribution exactly at the critical energy \cite{Santos2016,Santos2015}. This feature has important dynamical consequences.  

\subsection{Time-evolution after a quench}\label{sec:timeevolution}

\begin{center}
\begin{figure*}
\hspace*{-0.65cm}\includegraphics[width=1\textwidth]{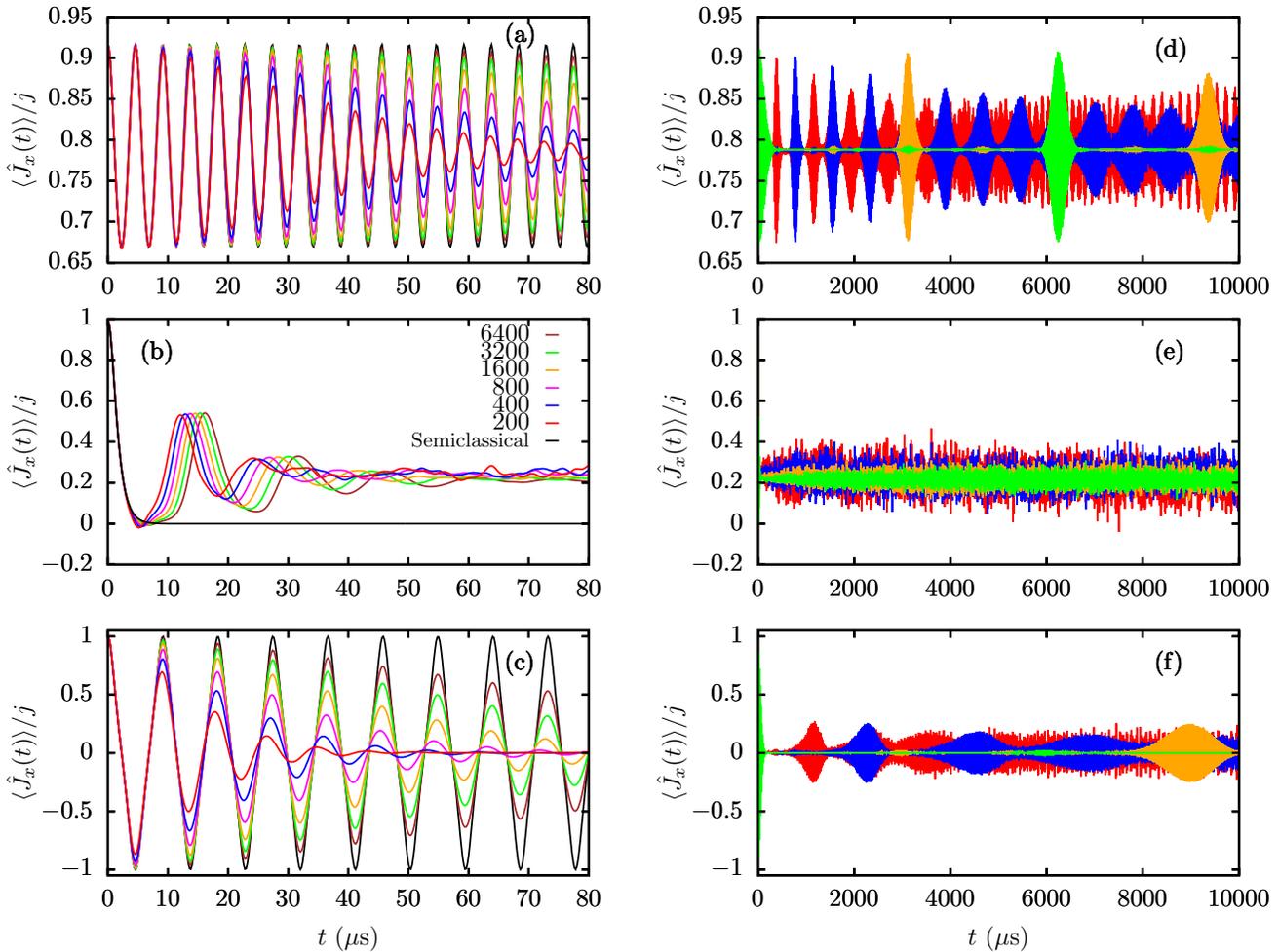}
\caption{Dynamics of the parity-breaking spin operator $\hat{J}_{x}$ after a quench $\lambda_{i}\to\lambda_{f}=1.75$. The initial state is of the form Eq. \eqref{eq:initialstate} with $\alpha=1/2$ and $\phi=0$, so the wavefunction is initially localized in the right energy well, $\langle \hatmath{C}\rangle=1$. The initial values of the control parameter and the average energy of the quench are (a,d) $\lambda_{i}=2.5$, $\epsilon(\lambda_{f})=-1.135$ (b,e) $\lambda_{i}=7$, $\epsilon(\lambda_{f})=-1$, and (c,f) $\lambda_{i}=27.5$, $\epsilon(\lambda_{f})=-0.89567$. The color code is indicated in (b) is followed in all panels. Semiclassical results are showed in (a,b,c) with black lines. }
\label{fig:paneljxtclassical}
\end{figure*}
\end{center}

Now we move on to study the dynamics of physical observables for each of the quenches in Fig. \ref{fig:ldosmeanspacing}. The abrupt change of the control parameter $\lambda_{i}\to\lambda_{f}$ forces the initial state out of equilibrium and the dynamics begin. In Fig. \ref{fig:paneljxtclassical} we focus on the quantum dynamics of the parity-breaking operator $\hat{J}_{x}$, whose expectation value at a given time $t$ in the time-evolving wavefunction $\ket{\Psi_{t}(\lambda_{f})}$ is

\beq \langle \hat{J}_{x}(t)\rangle=\bra{\Psi_{t}(\lambda_{f})}\hat{J}_{x}\ket{\Psi_{t}(\lambda_{f})}.
\eeq 
This operator plays an important role in DPTs-I as it can be used to define an order parameter \cite{Marino2022} [i.e., $m(t)=\langle \hat{J}_{x}(t)\rangle$ in Fig. \ref{fig:phasediagram}]. We also consider its classical counterpart: it is given by the dynamical function Eq. \eqref{eq:jxclassical}, and its evolution is completely determined by the Hamilton equations Eqs. (\ref{eq:partialP},\ref{eq:partialQ}). As the system size is increased, the quantum dynamics approaches the classical dynamics, which provides the exact evolution of the collective quantum system in the large-$j$ limit. In Fig. \ref{fig:paneljxtclassical}(a-c) an oscillatory pattern is observed in all cases. For smaller $j$, the quantum dynamics follows the classical expectation only for relatively short times, and for longer times we observe damping combined with dephasing with respect to the perfect $j\to\infty$ oscillation. The time when the different quantum results deviate from the classical evolution increases as $j$ increases, but it shows some peculiarities depending on the region of the spectrum where the initial state ends after the quench. For example, in Fig. \ref{fig:paneljxtclassical}(a) the quenched state only significantly populates states below the ESQPT, $\langle \hatmath{C}\rangle=+1$ is conserved and this means that the quenched state remains within the right classical well. This can be observed in $\langle \hat{J}_{x}(t)\rangle$, which is a positive quantity for all $t$. Likewise, classically, $j_{x}(t)\propto Q(t)$, which is positive in the right energy well [cf. Fig. \ref{fig:phasespace}(d)]. Fig. \ref{fig:paneljxtclassical}(c) represents an opposite scenario: the quench population is only significant above $E_{c}$, where $\langle \hatmath{C}\rangle$ is not conserved and the dynamics can explore both regions of the phase space. For this reason, $\langle \hat{J}_{x}(t)\rangle$ oscillates between positive and negative values, as can be understood again from the semiclassical picture, $j_{x}(t)\propto Q(t)$. An intermediate situation is considered in Fig. \ref{fig:paneljxtclassical}(b), where the average energy of the quench coincides with the ESQPT critical energy but both sides of the spectrum are nonetheless significantly populated. Classically, this trajectory corresponds to the curve `crossing itself' in Fig. \ref{fig:phasespace}(d). Starting from the initial state, the classical evolution $j_{x}(t)$ shows first a decay and then plateaus at zero, where the unstable fixed point trademark of the ESQPT takes place. Although a trajectory lying exactly on the critical line of the phase space has in principle access to the other side, the time that it takes to leave the fixed point diverges. The quantum dynamics, $\langle \hat{J}_{x}(t)\rangle$, shows drastic deviations from the classical expectation in this case.

In Fig. \ref{fig:paneljxtclassical}(d-f) we display the evolution of the same states as in \ref{fig:paneljxtclassical}(a-c) but for a longer time scale. Generally, after completely deviating from the classical prediction, the quantum dynamics oscillates around a steady-state value, and then it undergoes a dynamical revival that echoes its behavior at short times \cite{Milburn1997}. The evolution goes through a number of consecutive revivals until at very long times it eventually becomes very noisy with no clear pattern. This is clearly seen in Fig. \ref{fig:paneljxtclassical}(d,f). Both the time when the first revival occurs and the time interval between two consecutive revivals increase with system size. In Fig. \ref{fig:paneljxtclassical}(e) the situation is completely different: no such clear revivals are observed, and the dynamics simply fluctuates around the corresponding equilibrium value. 

\begin{center}
\begin{figure}[h]
\hspace*{-0.65cm}\includegraphics[width=0.52\textwidth]{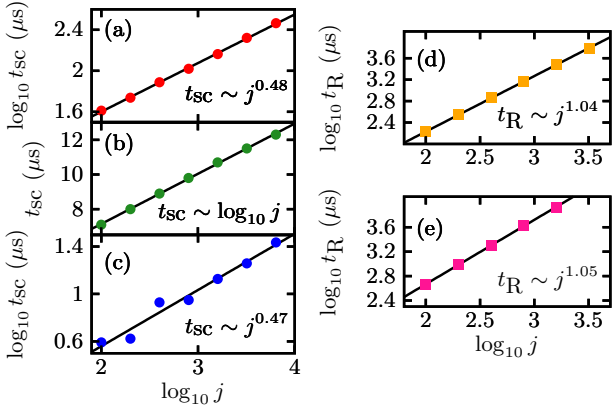}
\caption{Scaling of the semiclassical time $t_{\textrm{sc}}$ and the revival time $t_{\textrm{R}}$ as a function of the collective spin length $j$ for the time evolution of $\hat{J}_{x}$ from Fig. \ref{fig:paneljxtclassical}. (a-c) correspond to Fig. \ref{fig:paneljxtclassical}(a-c), respectively, while (d-e) correspond to Fig. \ref{fig:paneljxtclassical}(d) and Fig. \ref{fig:paneljxtclassical}(f), respectively.  }
\label{fig:paneltimes}
\end{figure}
\end{center}

Now we analyze two different time scales: the time when the quantum dynamics deviates from its large-$j$ semiclassical result, $t_{\textrm{SC}}$, and the time when the first revival occurs, $t_{\textrm{R}}$. To estimate $t_{\textrm{SC}}$, we consider the difference between quantum and classical results and compute the first value of $t$ for which this difference exceeds a given bound.  For the calculation of $t_{\textrm{SC}}$ this bound is $0.1$. Likewise, to estimate $t_{\textrm{R}}$, we compute the first time when the absolute value of the time evolution exceeds an arbitrary bound, only after the classical expectation has been completely lost. This bound is $0.85$ for Fig. \ref{fig:paneljxtclassical}(d) and 0.15 for Fig. \ref{fig:paneljxtclassical}(e). The results for the scaling with system size of these characteristic times are shown in Fig. \ref{fig:paneltimes}. On the one hand, for quenches with average energy below and above the ESQPT of Fig. \ref{fig:ldosmeanspacing}(a,c), that is $E<E_{c}$ or $E>E_{c}$, respectively, the time when the quantum dynamics deviates from the classical expectation follows a power-law behavior of the form $t_{\textrm{SC}}\sim \sqrt{j}$, as expected \cite{Marino2022,Lerose2019}; however, this time is much smaller for the quench ending at the ESQPT critical energy, revealing a logarithmic law instead, $t_{\textrm{SC}}\sim \log_{10} j$. Such a logarithmic scaling essentially precludes a realistic description of the quantum dynamics by means of the classical limit; for a macroscopic system with $N=10^{24}$ atoms, the quantum evolution would follow the semiclassical curve only up to $t\approx \log_{10} 10^{24}= 24$ $\mu$s, which is negligible compared to $t\approx \sqrt{10^{24}}=10^{12}$ $\mu$s $=10^{6}$ s as obtained for quenches ending below or above  $E_{c}$. For such macroscopic sizes, a dynamical phase transition takes place at the ESQPT: below or above the ESQPT, the dynamics shows persistent oscillations (possibly around a non-zero value in the first case, and around a zero value in the second case), while at the ESQPT no such scaling is possible and the dynamics simply fluctuates around a certain stationary value as in Fig. \ref{fig:paneljxtclassical}(b,e) after a extremely short time has elapsed. On the other hand, the first revival time is consistent with a simple linear behavior, $t_{\textrm{R}}\sim j$, with no revival taking place for quenches ending at the ESQPT.  

\subsection{Generalized microcanonical ensemble}\label{sec:gme}

If a closed quantum system reaches an equilibrium state, this state is necessarily equal to the infinite-time average of the real time-evolved wavefunction \cite{Reimann2008}. This means that long-time averages giving rise to dynamical order parameters for DPTs-I, regardless of whether they are real equilibrium states or just effective states around which the system oscillates, can be described with equilibrium ensembles. Hence, our first goal is to build an equilibrium ensemble, depending on all the relevant constants of motion, for an arbitrary system with the properties detailed in Sec. \ref{sec:setup}.

Consider a time-evolving wavefunction $\ket{\Psi_{t}}=\sum_{n}\sum_{k=\pm}c_{n,k}e^{-iE_{n,k}t}\ket{E_{n,k}}$. Above $E_{c}$, the long-time averaged density matrix,

\begin{equation}
\overline{\rho}=\lim_{\tau\to\infty}\frac{1}{\tau}\int_{0}^{\tau}\textrm{d}t\,\ket{\Psi_{t}}\bra{\Psi_{t}},
\end{equation}
only implies diagonal elements because all non-diagonal elements are removed by the time average,

\begin{equation}
\overline{\rho}(E>E_{c})=\sum_{n,k}|c_{n,k}|^{2}\ket{E_{n,k}}\bra{E_{n,k}},
\end{equation}
with $k=\pm$ labeling eigenstates of different parity. This result coincides with the diagonal ensemble \cite{Srednicki1999,Alessio2016}. However, if $E<E_{c}$, degeneracies in the spectrum imply that non-diagonal elements within the same eigenspace also survive: 

\begin{equation}\begin{split}
&\overline{\rho}(E<E_{c})=\sum_{n,k}|c_{n,k}|^{2}\ket{E_{n,k}}\bra{E_{n,k}}\\&+\sum_{n}c_{n,+}^{*}c_{n,-}\ket{E_{n,-}}\bra{E_{n,+}}+\sum_{n}c_{n,+}c_{n,-}^{*}\ket{E_{n,+}}\bra{E_{n,-}},
\end{split}\end{equation}
with $k=\pm$. Therefore, any statistical ensemble devised to describe equilibrium states below $E_{c}$ must include these non-diagonal elements. It is worth noting that these elements are complex-valued, and thus they cannot be described by a real non-diagonal operator like $\hatmath{C}$. For this reason, we need to define a second operator which is also constant below $E_{c}$, serving the purposes described above: 

\begin{equation}
\label{eq:K}
  \hat{{\mathcal K}} \equiv  \frac{i}{2} \left[ \hat{{\mathcal C}}, \hat{\Pi} \right].
\end{equation}
It is straightforward to show that $\hat{{\mathcal K}}$ commutes with
the energy projectors below the critical energy if $\hat{{\mathcal
    C}}$ does too. Indeed, from Eq. \eqref{eq:K} and using that $\hat{\Pi}\ket{E_{n\pm}}=\pm\ket{E_{n,\pm}}$ and $\hatmath{C}\ket{E_{n,\pm}}=\ket{E_{n,\mp}}$ (where the overall sign has been fixed to $+1$), one finds that $\hatmath{K}\ket{E_{n,\pm}}=\pm i \ket{E_{n,\mp}}$, evidencing the fact that $\hat{\Pi}$ and $\hatmath{K}$ cannot be diagonalized in the same eigenbasis as $\hatmath{K}$ flips the parity of any eigenstate $\ket{E_{n,\pm}}$ with $E_{n}<E_{c}$. Also, $[\hatmath{C},\hatmath{K}]\ket{E_{n,\pm}}=\pm 2i\ket{E_{n,\pm}}\neq 0$, meaning that $\hatmath{C}$ and $\hatmath{K}$ are not commuting either. Therefore, our generic system is characterized by a set of three non-commuting charges, $\lbrace \hat{\Pi}, \hat{{\mathcal C}}, \hat{{\mathcal K}} \rbrace$ for $E<E_c$. In an initial state of the form Eq. \eqref{eq:initialstate}, the expectation value of these three non-commuting charges can be evaluated exactly; the result is $\langle \hat{\Pi}\rangle=2\alpha-1$, $\langle \hatmath{C}\rangle=2\sqrt{\alpha(1-\alpha)}\cos\phi$, and $\langle \hatmath{K}\rangle=2\sqrt{\alpha(1-\alpha)}\sin\phi$.  However, only $\hat{\Pi}$ is a constant of motion for $E>E_{c}$. 
At first sight, this seems to imply that we need two different equilibrium ensembles to describe our generic system, and that the transient region between $E<E_{c}$ and $E>E_{c}$ is somehow ill-defined. However, we can fix this problem by defining two new operators to be used instead of $\hat{{\mathcal C}}$ and $\hat{{\mathcal K}}$,
\begin{eqnarray}
  \label{eq:defC}
  \widetilde{C} &=& \mathbb{I}_{E<E_c} \hat{{\mathcal C}} \, \mathbb{I}_{E<E_c}, \\
  \label{eq:defK}
  \widetilde{K} &=& \mathbb{I}_{E<E_c} \hat{{\mathcal K}} \, \mathbb{I}_{E<E_c}, 
\end{eqnarray}
where $\mathbb{I}_{E<E_c} \equiv \sum_{n} \theta_{n} \hat{P}_n$, $\hat{P}_n$
is the projector to the eigenspace with energy $E_n$, and
$\theta_n=1$ if $E_n<E_c$ and $\theta_n=0$ if
$E_n>E_c$. $\langle \widetilde{C}\rangle$ and $\langle \widetilde{K}\rangle$ are equal to $\langle \hat{{\mathcal C}}\rangle$ and $\langle \hat{{\mathcal K}\rangle}$ below the critical energy of the ESQPT, but identically zero above it. This means that $\widetilde{C}$ and $\widetilde{K}$ commute with the Hamiltonian in the TL, and therefore we can build an unique equilibrium ensemble from them. Furthermore, these two operators, together with $\hat{\Pi}$ and with the identity, close a SU(2) algebra in every subspace of degenerate energy levels. Indeed, within an energy subspace $\{\ket{E_{n,+}},\ket{E_{n,-}}\}$ these three operators can be written as follows: 

\beq
\hat{\Pi}=\mqty(1 & 0\\ 0 & -1),\,\,\widetilde{C}=\mqty(0 & 1 \\ 1 & 0),\,\,\widetilde{K}=\mqty(0 & -i\\ i & 0).
\eeq

Next, we use these operators to build a statistical ensemble for our setup. The basic idea is to rely on the set of non-commuting \cite{Guryanova2016,Halpern2016,Halpern2020} charges $\{\hat{\Pi},\widetilde{C},\widetilde{K}\}$ to reproduce any $2\times 2$ hermitian matrix accounting for all quantum coherences between parity sectors in a single energy eigenspace. The simplest choice for an such ensemble is 
\begin{equation}\label{eq:GME}
  \rho_{\textrm{GME}} (E, p, c, k) = \rho_\textrm{ME} (E) \left( \mathbb{I} + p \, \hat{\Pi} + c \, \widetilde{C} + k \, \widetilde{K} \right),
\end{equation}
where 
\beq
\rho_{\textrm{ME}}(E)=\frac{1}{2(N_{+}+N_{-})}\sum_{n}\left(\ket{E_{n,+}}\bra{E_{n,+}}+\ket{E_{n,-}}\bra{E_{n,-}}\right)
\eeq
denotes the standard microcanonical ensemble \cite{Alessio2016}, in which all
parity doublets, $\ket{E_{n,+}}$ and $\ket{E_{n,-}}$, within a small
energy window around the average energy value, $\langle E \rangle =
\textrm{Tr} [\hat{\rho} \hatmath{H}]$, are equally populated (irrespective of whether these parity doublets are degenerate), and
$\textrm{Tr} \rho_\textrm{GME}(E)=1$. Here, $N_{\pm}$ denotes the number of parity doublets above ($N_{+}$)/below ($N_{-}$) $E_{c}$ populated by the quench (see next subsection). This distinction is always possible because $\hat{\Pi}$ is an exact $\mathbb{Z}_{2}$ symmetry and thus the full Hamiltonian matrix can be directly split into a positive parity and negative parity Hamiltonian matrices. We call $\rho_{\textrm{GME}}(E,p,c,k)$ {\em
generalized microcanonical ensemble} (GME). Besides the average energy, it depends on three free parameters, $p,c,k\in\mathbb{R}$, which are fixed by requiring that $\textrm{Tr} [ \rho_{\textrm{GME}} \hat{\Pi}] = \langle \hat{\Pi} \rangle$, $\textrm{Tr}[ \rho_{\textrm{GME}} \widetilde{C}] = \langle \widetilde{C}\rangle$, and $\textrm{Tr}[ \rho_{\textrm{GME}} \widetilde{K}] = \langle \widetilde{K} \rangle$. Explicitly, in the large-$j$ limit these values can be worked out exactly,

\beq\label{eq:micropi}
\langle\hat{\Pi}\rangle=p,
\eeq

\beq\label{eq:microc}
\langle \widetilde{C}\rangle=c\frac{N_{-}}{N_{+}+N_{-}},
\eeq
and

\beq\label{eq:microk}
\langle\widetilde{K}\rangle=k\frac{N_{-}}{N_{+}+N_{-}},
\eeq
whence one may calculate the parameters $p,c,k$. 

This ensemble has the following properties: 

(i) It successfully accounts for the quantum coherences between parity sectors if and only if $E<E_c$. $\rho_{\textrm{GME}}(E,p,c,k)$ has off-diagonal elements in the parity eigenbasis if $c\neq 0$ and/or $k \neq 0$. As a consequence, the long-time averaged expectation value of parity-breaking observables, like $\textrm{Tr} \left[ \rho_{\textrm{GME}}(E,p,c,k) \hat{J}_x \right]$, may be different from zero only if $E<E_{c}$. Also, not every initial condition leads to a broken-symmetry equilibrium state if $E<E_{c}$; this only happens if $c\neq0$ and/or $k\neq 0$. Thus, it is possible for find non-zero order parameters of DPTs-I, i.e., there exists $\overline{m}\neq 0$ [cf. Fig. \ref{fig:phasediagram}].

(ii) It becomes diagonal when all populated states are above $E_c$. Hence, if $E>E_c$, $\textrm{Tr}\left[ \rho_{\textrm{GME}}(E,p,c,k) \hat{J}_x \right]=0$ for any initial condition, i.e., order parameters of DPTs-I are always $\overline{m}=0$ [cf. Fig. \ref{fig:phasediagram}].

These two points imply that {\em a DPT-I happens when a quench crosses the critical energy of the ESQPT}. If the initial state fulfills $E<E_c$ and the quench leads it to a spectrum region where all populated states are above $E_c$, then all information about quantum coherence between parity sectors is lost.  

Before moving on to the numerical results, a comment regarding the physical interpretation of the GME is in order. The operator $\hatmath{C}$ has a clear physical meaning. For an ensemble of classical particles described by Eq. \eqref{eq:initialstate}, it simply counts the number of particles within each disjoint part of the phase space, if $E<E_{c}$. However, neither $\hatmath{K}$ nor $\hat{\Pi}$ are admit such a classical interpretation. Quite contrarily, both these operators account for quantum correlations between the two disjoint classical regions: $\hat{\Pi}$ accounts for the real part of such correlations, and $\hatmath{K}$ for the imaginary part. Hence, the GME defined in Eq. \eqref{eq:GME} supposes that equilibrium states and long-time averages keep information about those quantum correlations, even in the TL.

As a side note, it is interesting to compare the statistical ensemble thus defined with the corresponding result that can be obtained using the ideas of the generalized Gibbs ensemble (GGE) \cite{Vidmar2016,Jaynes1957,Rigol2007}. Since the operators $\hat{\Pi}$, $\widetilde{C}$ and $\widetilde{K}$ all commute with $\hatmath{H}$ but are mutually non-commuting, the density matrix can be written $\rho_{\textrm{GGE}}(\beta_{p},\beta_{c},\beta_{k};\beta)=e^{-\beta \hatmath{H}-\beta_{p}\hat{\Pi}-\beta_{c}\widetilde{C}-\beta_{k}\widetilde{K}}$, where $\beta=1/k_{B}T$ is the inverse temperature, $k_{B}$ is the Boltzmann constant, and $\beta_{p}$, $\beta_{c}$ and $\beta_{k}$ are the multipliers associated to each charge. These parameters are fixed through the expectation values of the non-commuting charges in the GGE, i.e., $\langle \hatmath{O}\rangle=\beta_{O} \tanh({\beta_{p}^{2}+\beta_{c}^{2}+\beta_{k}^{2}})/\sqrt{\beta_{p}^{2}+\beta_{c}^{2}+\beta_{k}^{2}}$ where $\hatmath{O}$ can be $\hat{\Pi}$, $\widetilde{C}$ and $\widetilde{K}$ and $\beta_{O}=\beta_{p},\beta_{c},\beta_{k}$ accordingly. Thus, there is a direct correspondence between the results provided by the GME and by the GGE. However, there is a major practical drawback in the GGE in this case: When all expectation values vanish but one, e.g., when $\langle \hat{\Pi}\rangle=\langle \widetilde{K}\rangle=0$ and $\langle \widetilde{C}\rangle=1$, there is an infinite-valued temperature associated to the non-vanishing charge, $\beta_{c}\to\infty$ ($\beta_{p}=\beta_{k}=0$). This choice of parameters describes some of the states used in the literature, e.g. \cite{Muniz2020}. Since these parameters are usually obtained by fitting, working with the GME instead of the GGE is much more advantageous in this case.

\subsubsection*{Numerical results}
In computing the GME for a quench process we proceed as follows. We consider the eigenstates of the final Hamiltonian at $\lambda_{f}$ that the initial state at $\lambda_{i}$ populates, Eq. \eqref{eq:populationquench}, and calculate its average energy, $\langle E\rangle=\sum_{n}\sum_{k=\pm}|c_{n,k}|^{2}E_{n,k}(\lambda_{f})$. As in a standard microcanonical ensemble \cite{Alessio2016}, in the GME it is assumed that all states within a certain energy window $\Delta E$ centered at the average energy, $[\langle E\rangle-\Delta E,\langle E\rangle+\Delta E]$, are equally populated. 
The microcanonical energy window $\Delta E$ is composed of the $2N+1$ levels of positive parity around the target energy $\langle E\rangle$ and the $2N+1$ levels of negative parity. In our case, we have considered an energy window $\Delta E=2\sigma$ where $\sigma$ is the standard deviation of the distribution of populated states after the quench, i.e., $\sigma^{2}=\sum_{n}\sum_{k=\pm}|c_{n,k}|^{2}(E_{n,k}(\lambda_{f})-\langle E\rangle)^{2}$. We count the number of parity doublets (regardless of whether or not the corresponding energies $E_{n,\pm}$ are degenerate) below and above $E_{c}$, $N_{-}$ and $N_{+}$. Then making use of Eqs. (\ref{eq:micropi}, \ref{eq:microc}, \ref{eq:microk}), we obtain $p,c,k$.

On the one hand, for a state $E_{n,k}\leq E_{c}$ within the microcanonical window, the matrix form of the GME in the single energy subspace $\{\ket{E_{n,+}},\ket{E_{n,-}}\}$ is

\beq\label{eq:matrixbelow}
\rho_{n}(E_{n,k}\leq E_{c})=\frac{1}{2}\mqty(1+p & c-ik \\ c+ik & 1-p).
\eeq
Clearly, $\textrm{Tr}\,[\rho_{n}^{2}(E_{n,k}\leq E_{c})]=(1+p^{2}+c^{2}+k^{2})/2$, so any physical state must satisfy $p^{2}+c^{2}+k^{2}\leq 1$.
 
 On the other hand, for a state $E_{n,k}>E_{c}$ within the energy window, the GME takes the following form 

\beq\label{eq:matrixabove}
\rho_{n}(E_{n,k}>E_{c})=\frac{1}{2}\mqty(1+p & 0\\ 0 & 1-p).
\eeq
Note that Eq. \eqref{eq:matrixabove} is diagonal in the basis $\{\ket{E_{n,+}},\ket{E_{n,-}}\}$. For any other state $E_{n,k}\not\in [\langle E\rangle-\Delta E,\langle E\rangle+\Delta E]$, the corresponding matrix block is simply the null matrix, $\rho_{n}=0\times \mathbb{I}_{2}$. In the full parity basis $\{\ket{E_{1,+}},\ket{E_{1,+}},\ldots,\ket{E_{N,+}},\ket{E_{N,-}}\}$, the complete density matrix of the GME is a block-diagonal matrix containing each of the previous blocks, $\rho_{\textrm{GME}}=\textrm{diag}\,(\{\rho_{n}\}_{n})/Z$, where $Z=N_{+}+N_{-}$ is a normalization constant. Note that Eqs. (\ref{eq:matrixbelow},\ref{eq:matrixabove}) are always applicable, even when the average energy of the quench coincides with that of the ESQPT, $\langle E\rangle = E_{c}$; in such a case, the GME necessarily has contributions coming from states at both sides of the ESQPT. Therefore, the GME is built as explained above also in this case, with a density matrix containing these two contributions.

Once the GME has been built, one may compare the long-time average of a given physical observable $\hatmath{O}$, 
\beq\label{eq:longtimeaverage}
\overline{\langle \hatmath{O}\rangle}=\lim_{\tau\to\infty}\frac{1}{\tau}\int_{0}^{\tau}\textrm{d}t\,\bra{\Psi_{t}(\lambda_{f})}\hatmath{O}\ket{\Psi_{t}(\lambda_{f})},
\eeq
with the predictions of the GME, 
\beq\label{eq:gmeprediction}\langle \hatmath{O}\rangle_{\textrm{GME}}=\textrm{Tr}\,[\rho_{\textrm{GME}}\hatmath{O}].\eeq 

\begin{center}
\begin{figure}[h!]
\hspace*{-0.52cm}\includegraphics[width=0.53\textwidth]{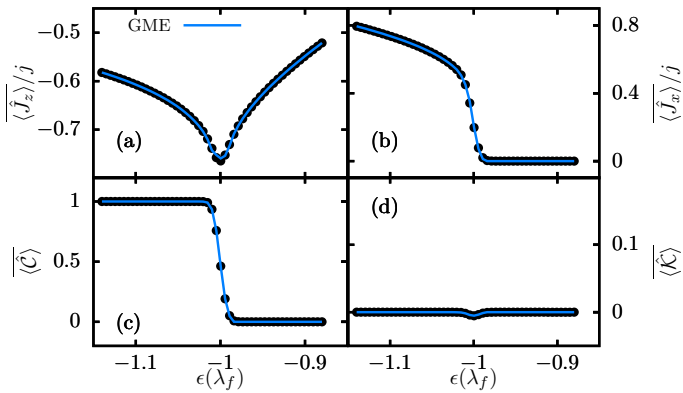}
\caption{(a-d) Long-time average, Eq. \eqref{eq:longtimeaverage}, of physical observables after a quench $\lambda_{i}\to\lambda_{f}=1.75$ as a function of the final energy, $\epsilon(\lambda_{f})$. System size is $j=6400$, and the initial state Eq. \eqref{eq:initialstate} has $\alpha=1/2$, $\phi=0$. Black points represent the exact averages while blue lines show the GME prediction, Eq. \eqref{eq:gmeprediction}. }
\label{fig:micros_largo}
\end{figure}
\end{center}

\begin{center}
\begin{figure}[h!]
\hspace*{-0.52cm}\includegraphics[width=0.5\textwidth]{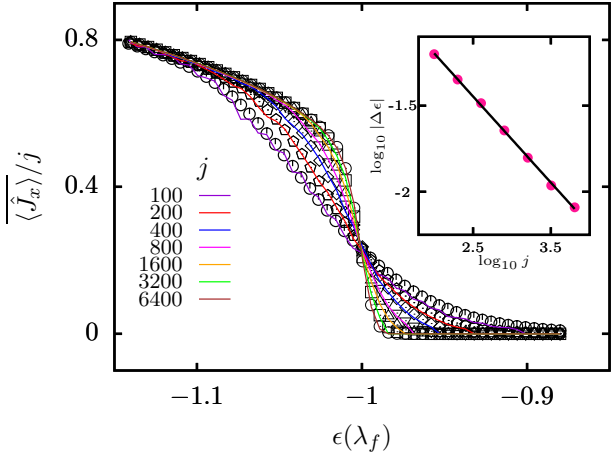}
\caption{Long-time average of $\hat{J}_{x}$ after a quench $\lambda_{i}\to\lambda_{f}=1.75$ as a function of the final energy $\epsilon(\lambda_{f})$. System sizes are indicated, and the initial state Eq. \eqref{eq:initialstate} has $\alpha=1/2$ and $\phi=0$. Black points represent the exact averages, while color lines show the GME prediction. Inset shows the scaling of $\Delta \epsilon=\epsilon_{c}(j)-\epsilon_{c}(\infty)$ with system size (bound $\gamma=1/20$), $|\Delta \epsilon|\sim 1/\sqrt{j}$.}
\label{fig:jxmicros}
\end{figure}
\end{center}

We have calculated the long-time average of relevant observables after a number of different quenches, letting the wavefunction relax in the final Hamiltonian during $10^{3}$ $\mu$s.  We always start in an initial state of the form Eq. \eqref{eq:initialstate} with $\alpha=1/2$, $\phi=0$ at different values of the control parameter $\lambda_{i}$, and these states are all quenched to $\lambda_{f}=1.75$. A case with $\phi\neq0$ is studied in \cite{Corps2022letter}. Depending on $\lambda_{i}$, the distribution of populated states in the final Hamiltonian is different, and thus so is the final mean energy $\epsilon(\lambda_{f})$ of the quench. These long-time averages are shown with points in Fig. \ref{fig:micros_largo} for $\hat{J}_{z}$, $\hat{J}_{x}$, $\hatmath{C}$ and $\hatmath{K}$ as a function of the final energy within the range $-1.1\lesssim \epsilon(\lambda_{f})\lesssim -0.9$, which goes through the ESQPT at $\epsilon_{c}=-1$. Irrespective of the observable considered, we can observe some precursors of non-analytic behavior around $\epsilon_{c}$. This non-analyticity is transferred directly from the level density to the expectation values of observables in systems with a single classical degree of freedom \cite{Cejnar2021}. Indeed, in Fig. \ref{fig:micros_largo}(a) we can see that the ESQPT critical point is signaled by an abrupt minimum in the long-time average of $\hat{J}_{z}$. As observed in Fig. \ref{fig:micros_largo}(b), $\hat{J}_{x}$ takes a non-zero value before the ESQPT has been crossed, and it vanishes once it has been completely crossed. In finite-$j$ systems, we observe a smooth transition between these scenarios, only becoming abrupt in the TL [cf. Fig. \ref{fig:jxmicros}]. In this sense, the long-time average of $\hat{J}_{x}$ can be considered as an order parameter of this DPT-I, $\overline{m}=\overline{\langle \hat{J}_{x}\rangle}$, occurring at the ESQPT, $\epsilon=\epsilon_{c}$, in the TL. This transition occurs because the phase delimited by $\epsilon<\epsilon_{c}$ is a broken-parity phase where the $\hat{\Pi}$ symmetry is broken, while it is restored right after crossing the ESQPT, $\epsilon>\epsilon_{c}$. In Fig. \ref{fig:micros_largo}(c, d) we focus on $\hatmath{C}$ and $\hatmath{K}$. Since the initial, broken-symmetry state Eq. \eqref{eq:initialstate} has $\alpha=1/2$ and $\phi=0$, its initial values for these operators are $\langle \hatmath{C}\rangle=1$ and $\langle\hatmath{K}\rangle=0$. The expectation value of $\hatmath{C}$ remains constant as long as $\epsilon\lesssim \epsilon_{c}$ \cite{Corps2022letter}, where this operator acts a constant of motion. For $\epsilon\gtrsim \epsilon_{c}$, it is no longer constant but oscillates, its average value vanishing completely. In the neighborhood of $\epsilon_{c}$ a smooth transition is again observed, which is a consequence of the finiteness of $j<\infty$. Finally, the long-time average of $\hatmath{K}$ is zero for $\epsilon<\epsilon_{c}$ due to the initial condition chosen, and it is also zero for $\epsilon>\epsilon_{c}$, when it is no longer constant. A case where $\overline{\langle\hatmath{K}\rangle}\neq 0$ for $\epsilon<\epsilon_{c}$ is discussed in \cite{Corps2022letter}. For all observables, the GME prediction has been depicted with a solid line. As can be seen, the agreement between the exact long-time averages and the GME is excellent in all cases. 

To end this section, we pay particular attention to the dynamical order parameter $\overline{\langle \hat{J}_{x}\rangle}$. The long-time averages have been represented for different values of $j$ approaching the TL in Fig. \ref{fig:jxmicros}. It is clearly observed that as $j$ increases, the transition from $\overline{\langle \hat{J}_{x}\rangle}\neq 0$ to $\overline{\langle\hat{J}_{x}\rangle}=0$ becomes sharper. The agreement with the GME, represented with color lines, improves as $j$ increases since the GME relies on constants of motion which become exact only in the TL. Importantly, all curves cross at some point around $\epsilon\approx \epsilon_{c}=-1$, suggesting a DPT-I in the TL, and this behavior is perfectly captured by the GME. To provide a definite answer, we estimate the precursor of the critical energy of the DPT-I, $\epsilon_{c}(j)$, by computing the last value of $\epsilon(\lambda_{f})$ for which $\overline{\langle \hat{J}_{x}\rangle}>\gamma$, where $\gamma$ is an arbitrary, small bound. Here, we choose $\gamma=1/20$. In the inset of Fig. \ref{fig:jxmicros} the difference between this precursor and the ESQPT critical energy in the TL, $\epsilon_{c}(\infty)=-1$, $|\Delta \epsilon|=|\epsilon_{c}(j)-\epsilon_{c}(\infty)|$, is represented as a function of the system size. This reveals a clear power-law behavior of the form $|\Delta \epsilon|\sim 1/\sqrt{j}$; to be precise, $|\Delta \epsilon|\sim 10^{a}j^{b}$ with $a=-0.195664$ and $b=-0.500194$. This is completely consistent with the DPT-I being caused by the ESQPT in the TL.

\subsection{Time-dependent protocol}\label{sec:timedependent}

In the previous section a quench was performed from an initial value of the coupling parameter, $\lambda_{i}$, to final value, $\lambda_{f}$, and the state was left to evolve at $\lambda_{f}$, which was held fixed. In this section we study the dynamics of the state subjected to an time-dependent slow process after a quench \cite{Puebla2015}. The protocol has the following steps.

(i) We prepare an initial state of the form Eq. \eqref{eq:initialstate} at $\lambda_{i}>\lambda_{c}$. 

(ii) We quench the state, $\lambda_{i}\to\lambda_{f}$. 

(iii) Immediately after performing the quench, we  solve the time-dependent Schr\"{o}dinger equation, $i\frac{\textrm{d}\ket{\Psi(\lambda(t))}}{\textrm{d}t}=\hatmath{H}(\lambda(t))\ket{\Psi(\lambda(t))}$, with $\lambda(t=0)=\lambda_{f}$, in order to implement a process where $\lambda(t)$ is a slowly-varying function\footnote{The goal of step (ii) is to excite the initial state so that the subsequent slow evolution of step (iii) drives it through the ESQPT. If step (ii) is skipped, then at most one may be able to drive the initial wavefunction through the QPT, which is not our focus. }.

At time $t$, the wavefunction can be formally expanded in the $\hat{J}_{z}$ eigenbasis $\{\ket{m}\}_{m=-j}^{j}$ as $\ket{\Psi_{t}(\lambda(t))}=\sum_{m=-j}^{j}\varphi_{m}(t)\ket{m}$. Thus, for the slow process we numerically solve the system of $2j+1$ coupled differential equations 

\begin{widetext}
\begin{equation}\label{eq:schrodinger}
\begin{split}
& i\frac{\textrm{d}}{\textrm{d}t}\varphi_{m}(t)=\varphi_{m}(t)\left[hm-\frac{\lambda(t)}{2N}(j(j+1)-m^{2})\right]\\&-\frac{\lambda(t)}{4N}\left[\varphi_{m+2}(t)\sqrt{j(j+1)-m(m+1)}\sqrt{j(j+1)-(m+2)(m+1)}+\varphi_{m-2}(t)\sqrt{j(j+1)-m(m-1)}\sqrt{j(j+1)-(m-2)(m-1)}\right],
\end{split}
\end{equation}
\end{widetext}
for all $m=-j,...,j$, and where $N=2j$ is the number of spin-1/2 particles. This affords the solutions $\{\varphi_{m}(t)\}_{m=-j}^{j}$ at different times according to the control parameter $\lambda(t)$. Here, we implement a forward-backward process between $\lambda(t=0)=\lambda_{0}$ and $\lambda(t=\tau)=\lambda_{1}$, with $\tau$ the time duration of each step of the protocol (forward or backward). In our choice, the control parameter $\lambda(t)$ is taken as the following linear function of time: 

\begin{equation}\label{eq:lambdat}
    \lambda(t)=\begin{cases}
        \lambda_{0}+\Delta\lambda \frac{t}{\tau}, &  0\leq t\leq \tau\\
        2\lambda_{1}-\lambda_{0}-\Delta\lambda\frac{t}{\tau}, & \tau\leq t\leq 2\tau
        \end{cases}
\end{equation}
where $\Delta\lambda=\lambda_{1}-\lambda_{0}$. The value of $\tau$ determines the rapidity of each process (forward or backward), i.e., how slowly or fast $\lambda(t)$ changes. For a perfectly adiabatic process, $\tau\to\infty$; however, relatively large values of $\tau$ afford results close enough to adiabaticity. Here we choose $\tau=500$ $\mu$s, which we find suitable for our purposes. 

In our simulations, we prepare different initial states of the form Eq. \eqref{eq:initialstate} with $\lambda_{i}=3$ and $\alpha=3/4$, characterized by different values of $\phi$. These initial states are parity-broken, and depending on $\phi$ they can be fully localized within one of the two classical energy wells or in a superposition of both. Then, we perform a quench to $\lambda_{0}\equiv \lambda_{f}=1.75$, and solve the time-dependent Schr\"{o}dinger equation from $\lambda_{0}$ to $\lambda_{1}=0.5$, and then from $\lambda_{1}$ back to $\lambda_{0}$.

\begin{center}
\begin{figure}[h]
\hspace{-0.3cm}
\includegraphics[width=0.49\textwidth]{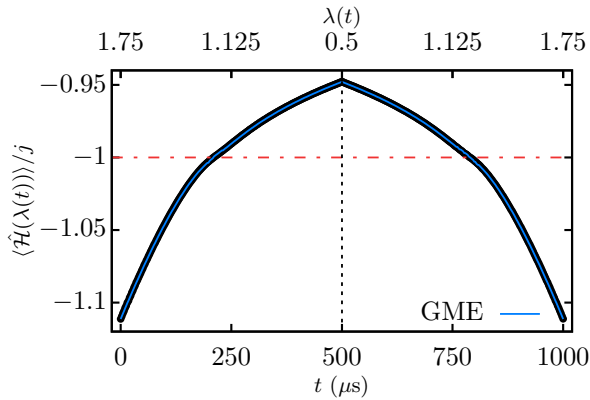} 
\caption{Energy across the slow process. The dashed vertical line separates the forward and backward steps. The blue line represents the GME result, and the horizontal red dotted-dashed line marks the ESQPT critical energy.  The system size is $j=1000$, and the driving parameter $\lambda(t)$ is taken as in Eq. \eqref{eq:lambdat}, with $\tau=500$. }
\label{fig:energyprocess}
\end{figure}
\end{center}

The energy of the state across the process is represented in Fig. \ref{fig:energyprocess}.  This figure clearly shows that the forward protocol drives the time-evolving wavefunction through the ESQPT at $\epsilon=-1$, and then crosses it back in the backward protocol. The curves are symmetric around $t=\tau$, which separates the forward and backward steps of the protocol. Note that no energy is `dissipated' during the entire process, i.e., the average energy of the initial and final states coincide: $\bra{\Psi(0)}\hatmath{H}(\lambda(0))\ket{\Psi(0)}=\bra{\Psi(2\tau)}\hatmath{H}(\lambda(2\tau))\ket{\Psi(2\tau)}$. We also plot the GME expectation for the process with a blue line.

\begin{center}
\begin{figure*}[t]
\hspace{-0.3cm}
\begin{tabular}{c c}
\includegraphics[width=0.5\textwidth]{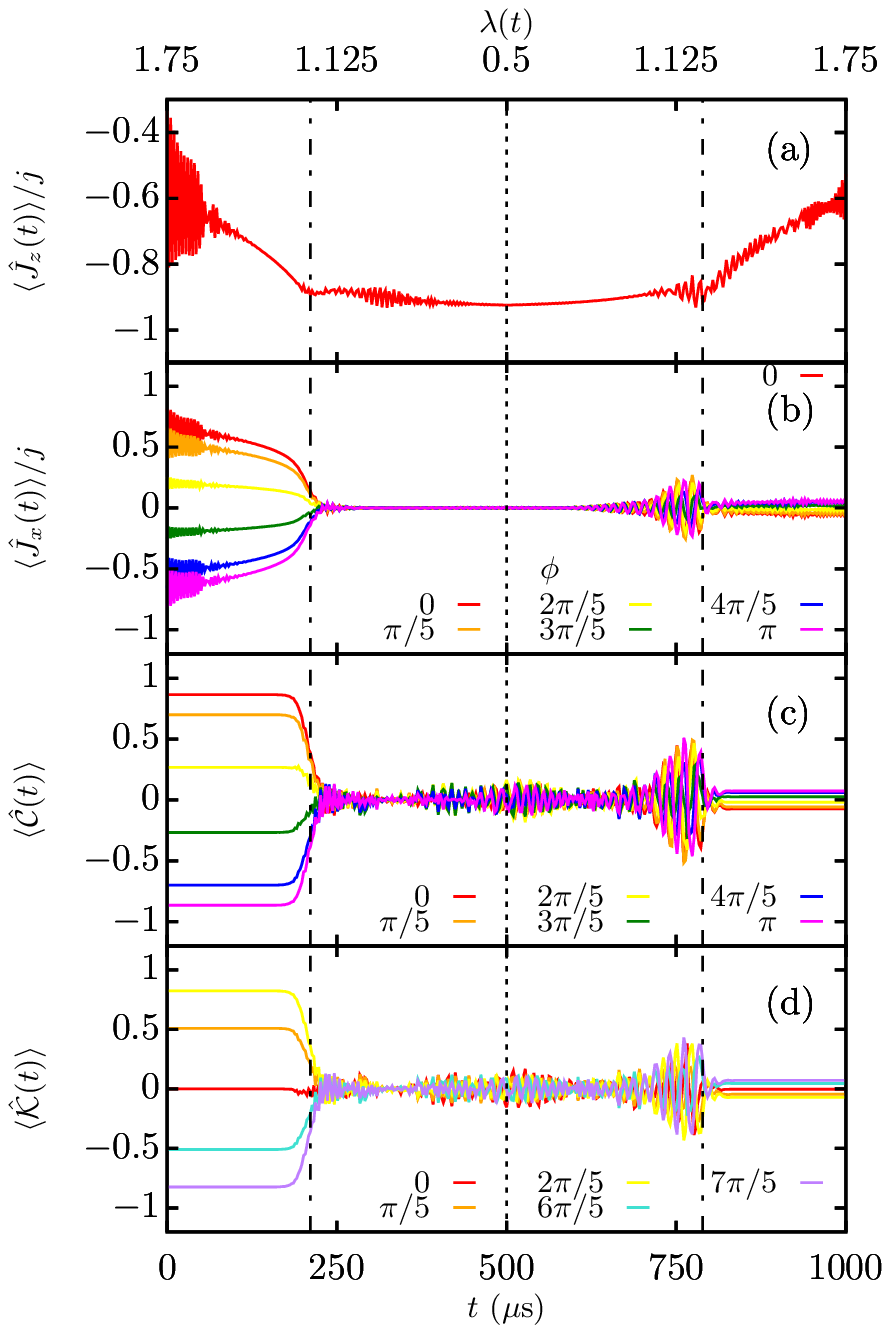} \includegraphics[width=0.5\textwidth]{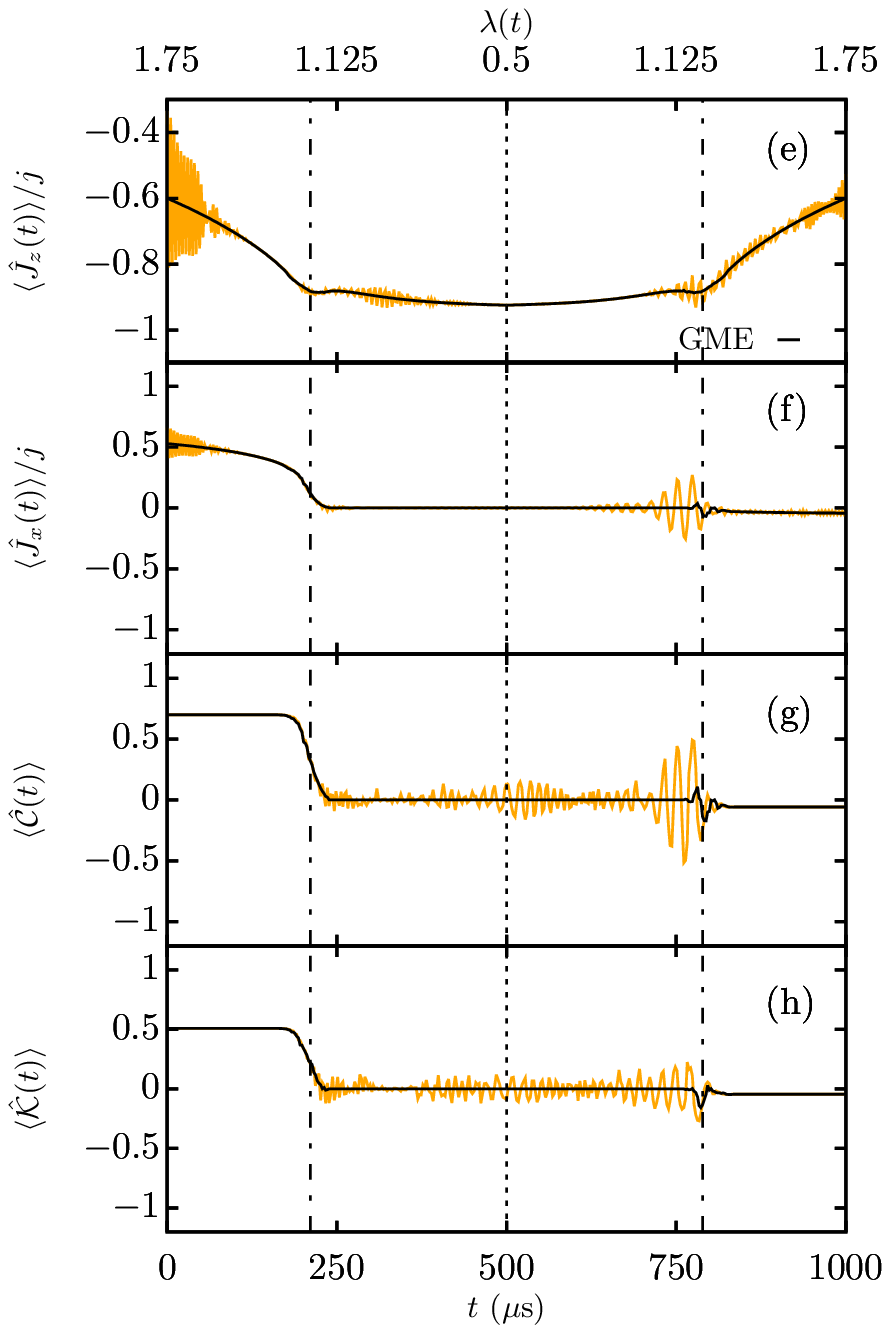}
\end{tabular}
\caption{(a-h) Time evolution after a quench $\lambda_{i}=3\to\lambda_{f}=1.75$ from a initial state of the form Eq. \eqref{eq:initialstate} with $\alpha=3/4$ and several $\phi$ values. System size is $j=1000$. The time values correspond to the expectation values of the corresponding observables in the time-evolving wavefunction as obtained from the time-dependent Schr\"{o}dinger equation Eq. \eqref{eq:schrodinger} with time-dependent $\lambda(t)$ Eq. \eqref{eq:lambdat}. The duration of the forward and backward processes is $\tau=500$ each. The vertical dashed line separates the forward and backward steps of the protocol, while the vertical dotted-dashed line depicts the time instant when the time-evolving wavefunction adiabatically crosses the ESQPT. In (e-h) results correspond to $\alpha=3/4$ and $\phi=\pi/5$, and the GME prediction is represented with a black dashed line. }
\label{fig:process}
\end{figure*}
\end{center}

As the wavefunction evolves in time, we monitor the evolution of representative observables. In Fig. \ref{fig:process}(a-d) we display such time expectation values from the initial states considered. A single curve for $\langle \hat{J}_{z}(t)\rangle$ is shown, corresponding to an initial state with $\phi=0$, because this observable does not depend on $\hatmath{C}$ or $\hatmath{K}$, and thus once $\alpha$ is fixed in Eq. \eqref{eq:initialstate} the time evolution is the same regardless of $\phi$. This time evolution is compatible with a reversible process because, oscillations aside, the expectation values at the beginning and at the end of the process are the same. This is in stark contrast with panels (b-d), where we focus on $\hat{J}_{x}$, $\hatmath{C}$, and $\hatmath{K}$. These observables are dependent on the population of classical energy wells and on the quantum correlations between them, given by $\phi$. Therefore, initial states with different $\phi$ yield different trajectories. Once $\alpha$ is fixed, the initial values of $\hat{J}_{x}$, $\hatmath{C}$ and $\hatmath{K}$ are related; in particular, $\langle \hatmath{C}\rangle \propto \cos\phi$ while $\langle\hatmath{K}\rangle\propto\sin\phi$. Despite these differences in the initial value of the observables, it is remarkable that all of them either become zero ($\hat{J}_{x}$) or oscillate around zero ($\hatmath{C}$, $\hatmath{K}$) after a certain time. This occurs around the same time when the average energy of the state is crossing the ESQPT, as shown in Fig. \ref{fig:energyprocess}, which is indicative of a DPT-I. After the forward protocol ends at $t=\tau$, the backward protocol begins and drives the system until $t=2\tau$. Even though the energy of the state at $t=0$ and $t=2\tau$ are the same [cf. Fig. \ref{fig:energyprocess}], the initial and final values of $\hat{J}_{x}$, $\hatmath{C}$, and $\hatmath{K}$ do not agree at all. This is a signature of the \textit{irreversibility} of the process: information about the initial state has been lost as a consequence of crossing the ESQPT. Indeed, when the ESQPT is crossed during the adiabatic process, the equilibrium density matrix $\hat{\rho}_{\textrm{eq}}=\lim_{t\to\infty}\frac{1}{t}\int_{0}^{t}\textrm{d}t'\,\hat{\rho}(t')$, with $\hat{\rho}(t)=\ket{\Psi(t)}\bra{\Psi(t)}$, becomes diagonal. Since the driving time $\tau$ in the protocol is larger than the semiclassical time $t_{\textrm{SC}}$ related to diffusion of the wavepacket,  $\hat{\rho}(t)$ is always close to $\hat{\rho}_{\textrm{eq}}$. Once the information contained in the off-diagonal elements of $\hat{\rho}_{\textrm{eq}}$ has been erased at the ESQPT during the forward process, there is no way to recover this information when crossing back the ESQPT during the backward process. Since $[\hat{\Pi},\hatmath{H}(\lambda)]=0$, $\forall \lambda$, this information erasing mechanism does not affect the expectation values of $\hat{\Pi}$; however, it does affect $\langle \hatmath{C}\rangle$, $\langle\hatmath{K}\rangle$ and, in general, all physical magnitudes whose equilibrium values depend on $\hatmath{C}$ and $\hatmath{K}$, such as $\hat{J}_{x}$ \cite{Corps2021}. This is clearly reflected in Table \ref{tab:values}, where we have collected the values of $p,c,k$ entering the GME at the beginning ($t=0$) and at the end ($t=2\tau$) of the adiabatic process. While $p\propto \langle \hat{\Pi}\rangle$ remains constant, the values of $c\propto\langle\hatmath{C}\rangle$ and $k\propto\langle\hatmath{K}\rangle$ are completely different, these values at $t=0$ coinciding with the exact expectations for $\alpha=3/4$ and $\phi=\pi/5$. This confirms that the information of the initial state is not recovered in the backward step of the protocol. Such loss of information can be quantified further through the von Neumann entropy $S=-\Tr[\rho_{n}\log \rho_{n}]$ of each $2\times 2$ block of the GME. The initial, $S(t=0)$, and final, $S(t=2\tau)$, values of the von Neumann entropy are calculated in Table \ref{tab:values}. The growth of the entropy is a consequence of information erasing, which is quantified by the GME. The DPT-I is therefore linked to a mechanism for information erasing, where the details of the initial condition are lost. For a detailed discussion of this mechanism, see Ref. \cite{Puebla2015}. 

\begin{table}[h!]
\begin{center}
\setlength\extrarowheight{5pt}
 \begin{tabular}{|| c || c c c c ||} 
 \hline
 Time & \hspace{0.4cm} $p$ & \hspace{0.4cm} $c$ & \hspace{0.4cm} $k$ & \hspace{0.4cm} $S$ \\ [0.5ex] 
 \hline\hline
$t=0$ & \hspace{0.4cm} $0.5$  & \hspace{0.4cm} $0.7006$ & \hspace{0.4cm} $0.5090$  & \hspace{0.4cm} $2.3\times 10^{-4}$ \\[1ex]
 \hline
 $t=2\tau$  & \hspace{0.4cm} $0.5$ & \hspace{0.4cm} $-0.0576$ & \hspace{0.4cm} $-0.0458$ & \hspace{0.4cm} $0.5594$ \\[1ex]
 \hline
\end{tabular}
\end{center}
\caption{Values of $p,c,k$ and the von Neumann entropy $S$ at the beginning ($t=0$) and end ($t=2\tau$) of the adiabatic process in Fig. \ref{fig:process}(e-h).}
\label{tab:values}
\end{table}

To end this section, we address the suitability of the GME to describe the time expectation values in this scenario. In our protocol the state is never allowed to relax before taking the expectation values of observables; this is to say that oscillations inherent of short-time dynamics are present in our results. Even though an equilibrium state may not be reached for short times, the GME is still appropriate to describe the \textit{average value} around which the wavefunction fluctuates, but not the exact form of the fluctuations themselves. In Fig. \ref{fig:process}(e-h) we focus on a single initial state with $\alpha=3/4$ and $\phi=\pi/5$, and represent the same time evolution as in Fig. \ref{fig:process}(a-d). On top of the numerical results, the GME expectation is represented with a black line. The agreement is excellent.

\section{DPT-II: Return probability}\label{sec:dptii}

In general, DPTs-II have a completely different origin than DPTs-I. Connections between both kinds of DPTs seem to exist in several systems of different nature \cite{Zunkovic2018,Lang2018,Weidinger2017,Hashizume2022,Sehrawat2021,Zunkovic2015,Lerose2019,Lang2018concurrence}, e.g. the non-analytical points of DPTs-II have been linked with zeros of the order parameters of DPTs-I. However, a common triggering mechanism for DPTs-II remains elusive. To study DPTs-II in the collective systems of our interest, we start by reviewing some common concepts.

In the seminal paper Ref. \cite{Heyl2013}, a new kind of phase transition was identified, occurring at some so-called \textit{critical times} when the overlap amplitude 
\beq\label{eq:G}
G(t)=\bra{\Psi_{0}(\lambda_{i})}e^{-i\hatmath{H}(\lambda_{f})t}\ket{\Psi_{0}(\lambda_{i})}
\eeq 
of a given initial state, $\ket{\Psi_{t=0}(\lambda_{i})}$, with itself after unitary evolution following a quench, $\ket{\Psi_{t}(\lambda_{f})}=e^{-i\hatmath{H}(\lambda_{f})t}\ket{\Psi_{0}(\lambda_{i})}$, becomes non-analytic. Such critical times were analytically obtained in the paradigmatic one-dimensional transverse-field Ising model \cite{Sachdev1999} (i.e., with nearest-neighbor couplings, rather than the fully connected model Eq. \eqref{eq:lipkin}). Viewed as a function on the complex plane [$z=it$, $G(t)=\bra{\Psi_{0}(\lambda_{i}))}e^{-z\hatmath{H}}\ket{\Psi_{0}(\lambda_{i})}$], Eq. \eqref{eq:G} bears a formal resemblance to partition functions in standard statistical mechanics, $\mathcal{Z}_{\beta}=\textrm{Tr}\,e^{-\beta\hatmath{H}}$. Since the free energy per particle $F=-(1/\beta N)\ln \mathcal{Z}_{\beta}$ becomes non-analytic in equilibrium phase transitions, an analogy can be drawn whereby the intensive, time-dependent quantity $g(t)=-(1/N)\ln G(t)$, known as \textit{rate function}, may signal a dynamical phase transition at certain critical times. Here, $N$ is the number of degrees of freedom of the system. We note that the square of the overlap Eq. \eqref{eq:G} is simply the survival probability, $\textrm{SP}(t)\equiv |G(t)|^{2}$, which we will consider later. For systems with broken-symmetry phases, non-analytic times may be studied with a somewhat different quantity as done in Ref. \cite{Heyl2014} in the $XXZ$ chain (see below). This is the subject of interest for this paper as the fully connected Ising model Eq. \eqref{eq:lipkin} is a perfect example of broken-symmetry models. In particular, it is worth emphasizing that in Refs. \cite{Heyl2013,Heyl2014} the DPT-II appears in a quench protocol that crosses the QPT. In other, more recent works, especially in models with long-range interactions or in collective (infinite-range interaction) systems, two scenarios can be distinguished: (i) a DPT-II may appear after a quench to a critical value of the control parameter that does not coincide with the QPT value \cite{Sciolla2011}; this has been called a \textit{regular} dynamical phase; also, (ii) a DPT-II may also appear even if the QPT is not crossed, as observed in \cite{Homrighausen2017,Halimeh2017}; this has been associated with an \textit{anomalous} dynamical phase. This anomalous dynamical phase has been characterized by cusps appearing only \textit{after} the first minimum of the return rate (see below), while in the regular dynamical phase the first cusp appears always \textit{before} the first minimum \cite{Halimeh2017}.  

In \cite{Corps2022letter} we show that the regular dynamical phase is triggered by the ESQPT common of collective systems, and that the critical times only show a scaling typical of a phase transition in this phase (and not in the anomalous phase).  As in the previous sections, here we focus on the impact of the ESQPT on the appearance or absence of DPTs-II, delving deeper into these questions. We provide analytical results, valid in the TL, showing that the main mechanism \cite{Heyl2014} for DPTs-II is only allowed when the energy of the quenched state is above the critical energy of the ESQPT, $E>E_{c}$, while it is forbidden if $E<E_{c}$. We show that for $E>E_{c}$ one finds a regular dynamical phase of DPTs-II, while for $E<E_{c}$ this becomes an anomalous dynamical phase; thus, the change from regular to anomalous phase is triggered precisely by the ESQPT.

\subsection{Analytical results}\label{sec:analyticalargument}

To study DPTs-II, one considers as an initial state $\ket{\Psi_{0}(\lambda_{i})}$ a general superposition of the degenerate ground state in the degenerate phase (where the $\mathbb{Z}_{2}$ symmetry may be broken), in our case $E<E_{c}$. Then, a quench $\lambda_{i}>\lambda_{c}\to\lambda_{f}$ is performed, and the state is allowed to evolve in time under the new Hamiltonian, $\ket{\Psi_{t}(\lambda_{f})}$.  As mentioned before, in the case of broken-symmetry models DPTs-II are defined through the non-analytic times in the parity-projected return probability (PPRP) \cite{Heyl2014},

\beq\label{eq:echo}\begin{split}
\mathcal{L} (t)&=\left |\bra{E_{0,+}(\lambda_{i})}e^{-i \hatmath{H}(\lambda_{f}) t}\ket{\Psi_{0}(\lambda_{i})}\right|^{2}\\&+\left|\bra{E_{0,-}(\lambda_{i})}e^{-i \hatmath{H}(\lambda_{f}) t}\ket{\Psi_{0}(\lambda_{i})}\right|^{2}.
\end{split}
\eeq
For convenience, we may define

\beq
\mathcal{L}_{\pm}(t)=\left|\bra{E_{0,\pm}(\lambda_{i})}e^{-i\hatmath{H}(\lambda_{f})t}\ket{\Psi_{0}(\lambda_{i})}\right|^{2}
\eeq
so that $\mathcal{L}(t)=\mathcal{L}_{+}(t)+\mathcal{L}_{-}(t)$. Here, $\mathcal{L}_{\pm}(t)$ are the return probabilities to the positive-parity and negative-parity projections of the initial state \cite{Heyl2014} [note the difference with $G(t)$ in Eq. \eqref{eq:G}, where no projections to parity subspaces are considered]. Specifically, a DPT-II occurs at $t=t_{*}$ if $\mathcal{L}(t)$ is a non-analytic function at $t=t_{*}$. Instead of Eq. \eqref{eq:echo}, it is common to analyze the rate function

\beq\label{eq:erre}
r_{N}(t)=-\frac{1}{N}\ln \mathcal{L}(t),
\eeq
where $N$ is a TL parameter; in our case, $N=2j$. According to \cite{Heyl2014,Touchette2009}, each of the terms in the PPRP, $\mathcal{L}_{\pm}(t)$, follows a law

\beq\label{eq:scalingL}
\mathcal{L}_{\pm}(t)=e^{-N\Omega_{\pm}(t)},
\eeq
where $\Omega_{\pm}(t)$ is an intensive quantity. Therefore, 

\beq
r_{N}(t)=\Omega_{+}(t)-\frac{1}{N}\ln \left[1+e^{-N(\Omega_{-}(t)-\Omega_{+}(t))}\right]
\eeq
Regarding the second term on the right-hand side, in the TL (i) if $\Omega_{-}(t)>\Omega_{+}(t)$, then $\lim_{N\to\infty} -\frac{1}{N}\ln[1+e^{-N(\Omega_{-}(t)-\Omega_{+}(t))}]=0$, while (ii) if $\Omega_{+}(t)>\Omega_{-}(t)$, then     $\lim_{N\to\infty} -\frac{1}{N}\ln[1+e^{-N(\Omega_{-}(t)-\Omega_{+}(t))}]=\lim_{N\to\infty} -\frac{1}{N}\ln e^{N(\Omega_{+}(t)-\Omega_{-}(t))}=\Omega_{-}(t)-\Omega_{+}(t)$. Therefore, in the TL one has

\beq\label{eq:erret}
r(t)\equiv \lim_{N\to\infty}r_{N}(t)=\begin{cases}
\Omega_{+}(t), & \Omega_{-}(t)>\Omega_{+}(t)\\
\Omega_{-}(t), & \Omega_{-}(t)<\Omega_{+}(t).
\end{cases}
\eeq

Hence, there exists a singular point at $t=t_{*}$ when the functions $\Omega_{\pm}(t)$ intersect,  $\Omega_{+}(t_{*})=\Omega_{-}(t_{*})$. As in any other phase transition, it is therefore expected that $r_{N}(t)$ remains analytic at $t=t^{*}$ in finite-$N$ systems, and only becomes singular in the TL. From Eq. \eqref{eq:erret} it is obvious that the $n$th derivative of $r(t)$ is $\textrm{d}^{n}r(t)/\textrm{d}t^{n}=\textrm{d}^{n}\Omega_{\min}(t)/\textrm{d}t^{n}$ where $\Omega_{\min}(t)\equiv \min\{\Omega_{+}(t),\Omega_{-}(t)\}$. Thus, one may assign an \textit{order} to a DPT-II by considering the value of $n$ for which $\textrm{d}^{n}r(t)/\textrm{d}t^{n}$ becomes discontinuous. Note, however, that this result does not preclude the existence of other non-analytical points where $\Omega_{+}(t)$ and/or $\Omega_{-}(t)$ become non-analytic. We will come back this point later on.

To derive a theory for this kind of DPTs, we start by considering an initial state given by Eq. \eqref{eq:initialstate} at $\lambda_i$ quenched to $\lambda_f$. Since $\hat{\Pi}$ is an exact conserved quantity, we can expand the initial eigenvectors of $\hatmath{H}(\lambda_{i})$ as a combination of the final eigenvectors of $\hatmath{H}(\lambda_{f})$ \textit{of the same parity}. For example, the broken-symmetry ground state at $\lambda_{i}$ may be written

\beq\label{eq:expansion}
\ket{E_{0,\pm}(\lambda_{i})}=\sum_{n}c_{n,\pm}\ket{E_{n,\pm}(\lambda_{f})}.
\eeq
Therefore, we may rewrite the quenched state as a combination of eigenstates of the final Hamiltonian as 

\beq\begin{split}
\ket{\Psi_{t}(\lambda_{f})}&=\sqrt{\alpha}\sum_{n}c_{n,+}e^{-iE_{n,+}(\lambda_{f})t}\ket{E_{n,+}(\lambda_{f})}\\&+e^{i\phi}\sqrt{1-\alpha}\sum_{n}c_{n,-}e^{-i E_{n,-}(\lambda_{f})t}\ket{E_{n,-}(\lambda_{f})}.
\end{split}
\eeq
Now, parity conservation allows to write the terms of the PPRP as

\begin{widetext}
\beq
\mathcal{L}_{+}(t)=\left|\bra{E_{0,+}(\lambda_{i})}\ket{\Psi_{t}(\lambda_{f})}\right|^{2}=\left|\sqrt{\alpha}\sum_{n}c_{n,+}e^{-i E_{n,+}(\lambda_{f})t}\bra{E_{0,+}(\lambda_{i})}\ket{E_{n,+}(\lambda_{f})}\right|^{2}=\alpha\left|\sum_{n}|c_{n,+}|^{2}e^{-iE_{n,+}(\lambda_{f})t}\right|^{2},
\eeq
\end{widetext}
and, similarly,

\beq
\mathcal{L}_{-}(t)=(1-\alpha)\left|\sum_{n}|c_{n,-}|^{2}e^{-iE_{n,-}(\lambda_{f})t}\right|^{2}.
\eeq
Defining the complex-valued functions 
\beq\label{eq:fplusminus}
f_{\pm}(t)\equiv \sum_{n}|c_{n,\pm}|^{2}e^{-i E_{n,\pm}(\lambda_{f})t},
\eeq
the components of the PPRP are

\beq\label{eq:componentsL}
\mathcal{L}_{+}(t)=\alpha|f_{+}(t)|^{2},\,\,\,\,\,\,\mathcal{L}_{-}(t)=(1-\alpha)|f_{-}(t)|^{2}.
\eeq

The main result of this section is the following. Consequences will then follow.

{\bf Result:} If $E<E_c$, then $f_+(t) = f_-(t)$, $\forall t$.

Let us prove this result. First, since in the TL $E_{n,+}(\lambda_{f})=E_{n,-}(\lambda_{f})$ for all $n$ such that $E_{n,\pm}<E_{c}$ \cite{Corps2021}, the oscillatory parts in $f_{\pm}(t)$ are the same. Therefore, we only need to analyze the coefficients $c_{n,\pm}$. 

Because $\hat{\Pi}$ is an exact conserved quantity, $\bra{E_{n,+}(\lambda_{f})}\ket{E_{0,-}(\lambda_{i})}=0$ since these eigenstates belong to different parity sectors, and therefore

\beq
c_{n,+}=\bra{E_{n,+}(\lambda_{f})}\ket{E_{0,+}(\lambda_{i})}.
\eeq
For $E<E_{c}$, $\hat{\mathcal{C}}$ acts as a conserved quantity. It does not commute with parity because it changes the parity of any Fock state \cite{Corps2021}, 

\beq
\hat{\mathcal{C}}\ket{E_{0,\pm }(\lambda_{i})}=\ket{E_{0,\mp}(\lambda_{i})},
\eeq
where we have fixed the arbitrary overall sign to $+1$ \cite{Corps2021}. Also, the unitarity of $\hatmath{C}$ implies $\hatmath{C}^{\dagger}\hatmath{C}=1$. Therefore, 

\begin{equation}
\begin{split}
    \left| c_{n,+} \right| &= \left| \bra{E_{n,+} (\lambda_f)}\ket{E_{0,+} (\lambda_i)} \right|  \\ &=  \left| \bra{E_{n,-} (\lambda_f)}\hat{{\mathcal C}}^{\dagger} \hat{{\mathcal C}}\ket{E_{0,-} (\lambda_i)} \right|  \\ &=   \left| \bra{E_{n,-} (\lambda_f)}\ket{E_{0,-} (\lambda_i)} \right| = \left| c_{n,-} \right|.
    \end{split}
\end{equation}

 It follows that $f_+(t)=f_-(t)$ in the TL, if all the populated states are below the critical energy of the ESQPT. 
 
 This formal result has two immediate consequences.

{\bf Consequence 1:} the constancy of $\hat{{\mathcal C}}$ if $E<E_c$ implies $\Omega_+(t)$ and $\Omega_-(t)$ cannot intersect. Therefore {\em the mechanism for DPTs-II proposed in \cite{Heyl2014} is forbidden for quenches below the critical energy, $E<E_{c}$}. It is only allowed if the quench leads the state to $E>E_c$.

Let us assume that $f_{+}(t)=f_{-}(t)$ for all $t$; then, if $\alpha\in(0,1)$, Eq. \eqref{eq:componentsL} implies
\beq\label{eq:ratio}
\frac{\mathcal{L}_{+}(t)}{\mathcal{L}_{-}(t)}=\frac{\alpha}{1-\alpha}
\eeq
for all $t$ too. If $\alpha=1/2$, then $\mathcal{L}_{+}(t)=\mathcal{L}_{-}(t)$ for all $t$, and therefore it is clear that $\Omega_{+}(t)=\Omega_{-}(t)$ for all $t$, implying no crossing is possible. If $0<\alpha<1/2$, Eq. \eqref{eq:ratio} implies $\mathcal{L}_{+}(t)<\mathcal{L}_{-}(t)$ for all $t$, and from Eq. \eqref{eq:scalingL} this implies that $\Omega_{+}(t)>\Omega_{-}(t)$ for all $t$. Otherwise, $1/2<\alpha<1$, and in this case the contrary holds true, $\Omega_{-}(t)>\Omega_{+}(t)$. In neither case a crossing in $\Omega_{\pm}(t)$ is possible. 

Finally, in the trivial cases where $\alpha=1$ or $\alpha=0$, we have that either $\mathcal{L}_{+}(t)=|f(t)|^{2}\geq 0=\mathcal{L}_{-}(t)$ or $\mathcal{L}_{-}(t)=|f(t)|^{2}\geq  0=\mathcal{L}_{+}(t)$. Thus, no crossing is possible, as $\mathcal{L}_{+}(t)$ and $\mathcal{L}_{-}(t)$ are either equal or different for all time. 

This result sets an important bound on the region of the spectrum where the main mechanism for DPTs-II is not allowed to occur. 

Thus far we have focused on the indicators of DPTs-II in the return probability for $\mathbb{Z}_{2}$ broken-symmetry systems, Eq. \eqref{eq:echo}, proposed in Ref. \cite{Heyl2014}. Next we move onto the survival probability, closely related to the indicator used in Ref. \cite{Heyl2013}. The survival probability is also a measure of the overlap of the time-evolved state $\ket{\Psi_{t}(\lambda_{f})}$ with its initial value, $\ket{\Psi_{0}(\lambda_{i})}$, but no projections onto parity subspaces are considered:

\beq\label{eq:survivalprobability}
\textrm{SP}(t)=\left|\bra{\Psi_{0}(\lambda_{i})}\ket{\Psi_{t}(\lambda_{f})}\right|^{2}.
\eeq
As mentioned before, $\textrm{SP}(t)$ is simply the probability associated to the overlap $G(t)$, Eq. \eqref{eq:G}, considered in Ref. \cite{Heyl2013}. In principle, Eq. \eqref{eq:survivalprobability} is different from $\mathcal{L}(t)$ in Eq. \eqref{eq:echo}, because $\mathcal{L}(t)$ does not account for the interference between initial states of different parity. Considering an initial state of the form Eq. \eqref{eq:initialstate} at an initial coupling parameter $\lambda_{i}>\lambda_{c}$, and performing a quench $\lambda_{i}\to\lambda_{f}$, the survival probability Eq. \eqref{eq:survivalprobability} reads

\beq\begin{split}
\textrm{SP}(t)&=\Big|\alpha\sum_{n}c_{n,+}e^{-iE_{n,+}(\lambda_{f})t}\bra{E_{0,+}(\lambda_{i})}\ket{E_{n,+}(\lambda_{f})}\\&+(1-\alpha)\sum_{n}c_{n,-}e^{-iE_{n,-}(\lambda_{f})t}\bra{E_{0,-}(\lambda_{i})}\ket{E_{n,-}(\lambda_{f})}\Big|^{2}.
\end{split}
\eeq
Making use of Eq. \eqref{eq:expansion} and substituting in the definition of $f_{\pm}(t)$ in Eq. \eqref{eq:fplusminus}, this is 

\beq\label{eq:spf}
\textrm{SP}(t)=\left |\alpha f_{+}(t)+(1-\alpha)f_{-}(t)\right|^{2}.
\eeq
From this expression, we obtain two different behaviors depending on the energy of the state considered, which we make explicit below. 

(i) If $E<E_{c}$, then, according to our previous derivation, $f_{+}(t)=f_{-}(t)\equiv f(t)$ for all $t$, and therefore

\beq
\textrm{SP}(t)=|f(t)|^{2}=\mathcal{L}_{+}(t)+\mathcal{L}_{-}(t)=\mathcal{L}(t).
\eeq
This shows that, in the TL, Eq. \eqref{eq:echo} and Eq. \eqref{eq:survivalprobability} are \textit{equal} if $E<E_{c}$.  

(ii) If $E>E_{c}$ instead, then we have that $f_{-}(t)\neq f_{+}(t)$ in general. Thus,

\beq\begin{split}
\textrm{SP}(t)&=\alpha^{2}|f_{+}(t)|^{2}+(1-\alpha)^{2}|f_{-}(t)|^{2}\\&+\alpha(1-\alpha)\left[f_{+}(t)f_{-}^{*}(t)+f_{+}^{*}(t)f_{-}(t)\right]\\&\neq \alpha|f_{+}(t)|^{2}+(1-\alpha)|f_{-}(t)|^{2}=\mathcal{L}(t),
\end{split}
\eeq
implying that in this case Eq. \eqref{eq:echo} and Eq. \eqref{eq:survivalprobability} are different quantities. Obviously, if $\alpha=0$ or $\alpha=1$, then $\textrm{SP}(t)=\mathcal{L}(t)$, as in this case the initial state Eq. \eqref{eq:initialstate} has either positive or negative parity, and thus no interference is possible between different parity sectors. 

The second main consequence of the analytical results in this section is therefore the following: 

\textbf{Consequence 2:} If $E<E_{c}$, $\hatmath{C}$ acts a constant of motion in the TL and, thus, the return probability $\mathcal{L}(t)$, Eq. \eqref{eq:echo}, and the survival probability, $\textrm{SP}(t)$, Eq. \eqref{eq:survivalprobability}, coincide in the TL. If $E>E_{c}$, these quantities are, in general, different. 

In what follows we will see that this result is important in understanding the so-called anomalous DPT-II phase \cite{Homrighausen2017}. 

\subsection{Numerical results}\label{sec:numericalresults}

\begin{center}
\begin{figure*}[h]
\hspace{-0.3cm}
\begin{tabular}{c c}
\includegraphics[width=0.45\textwidth]{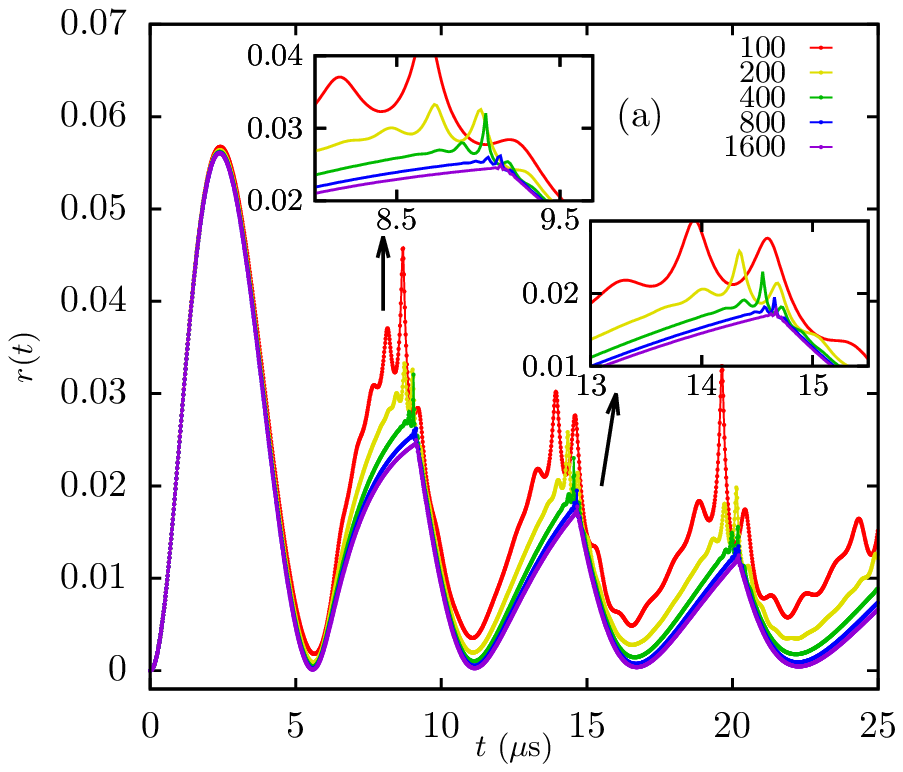} & \includegraphics[width=0.45\textwidth]{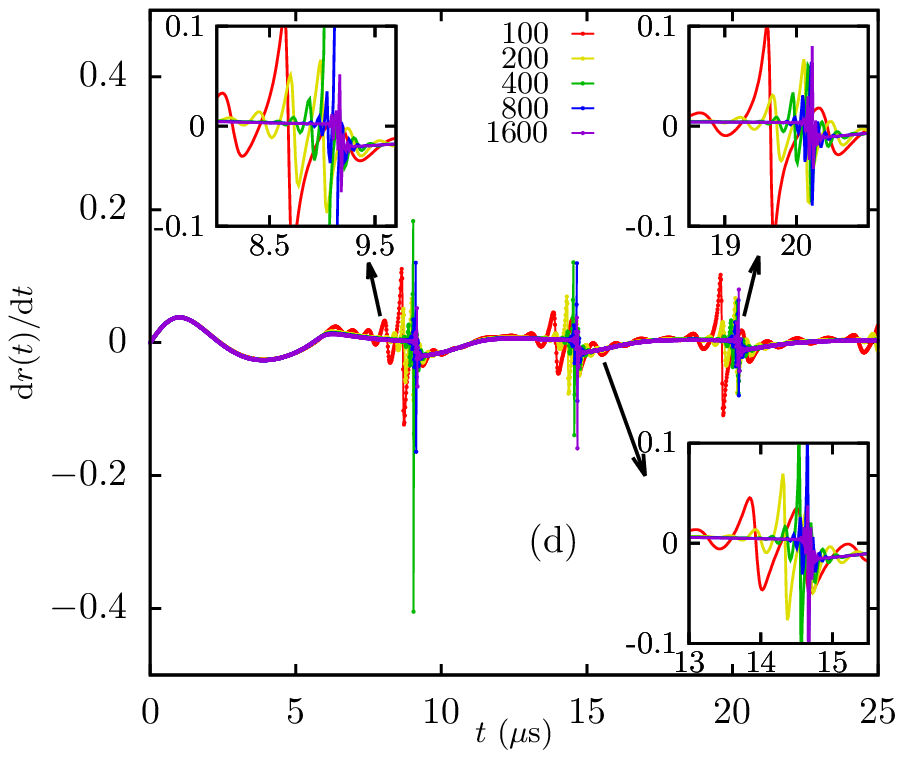} \\ \includegraphics[width=0.45\textwidth]{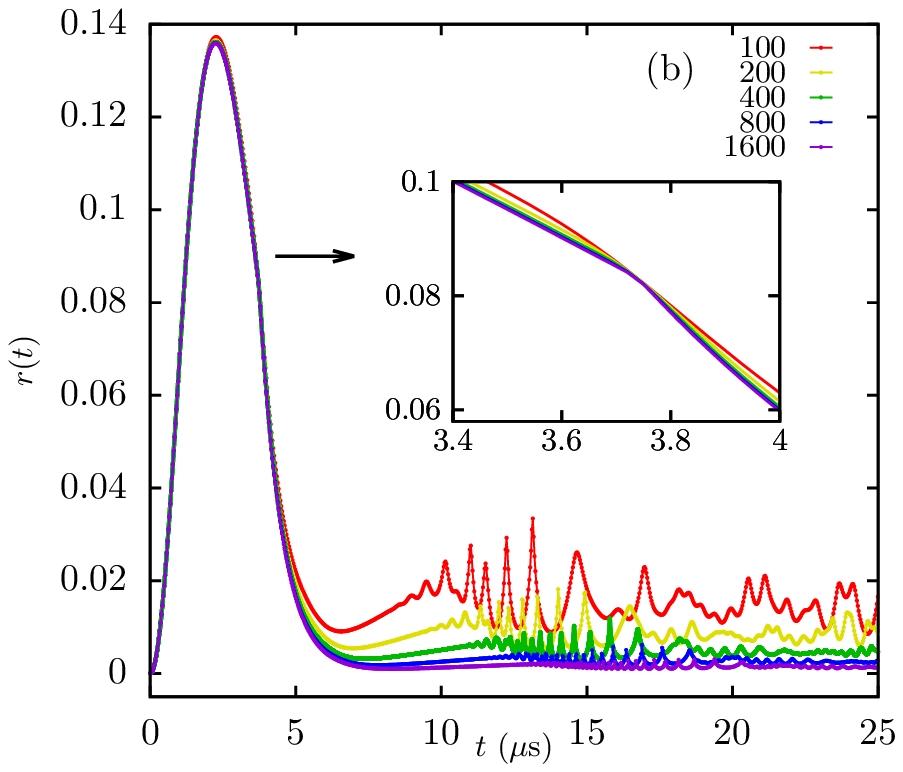} & \includegraphics[width=0.45\textwidth]{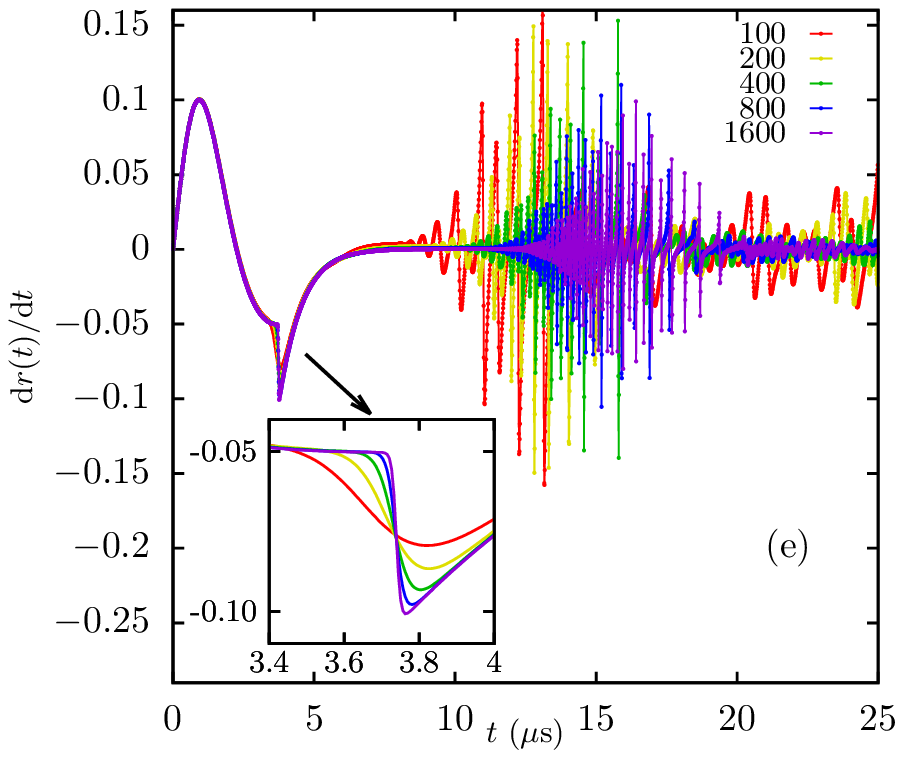} \\ \includegraphics[width=0.45\textwidth]{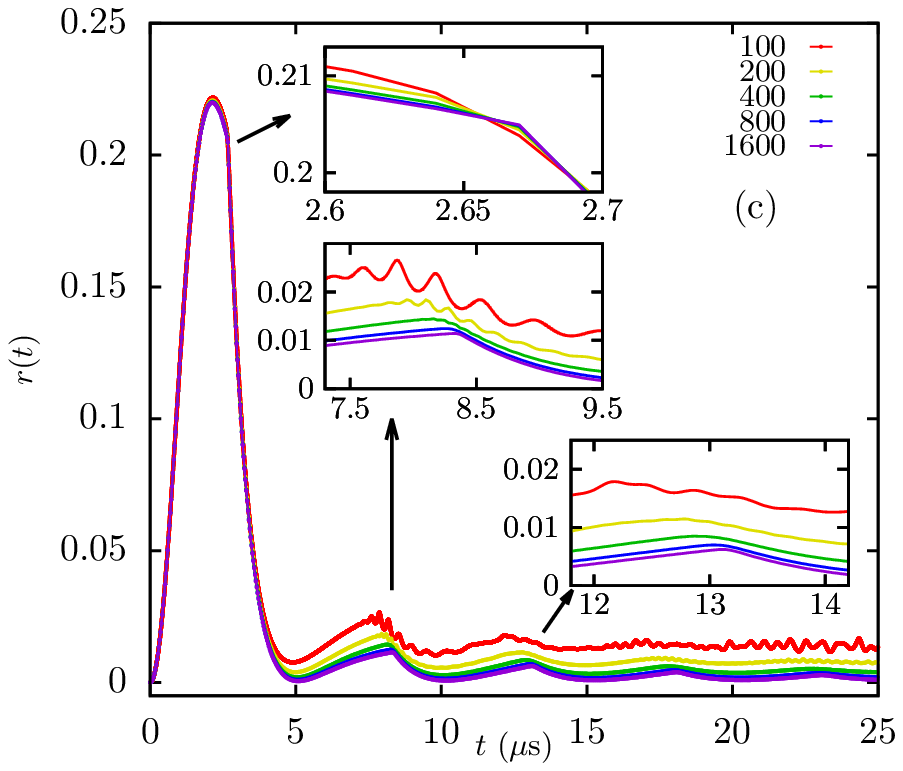} & \includegraphics[width=0.45\textwidth]{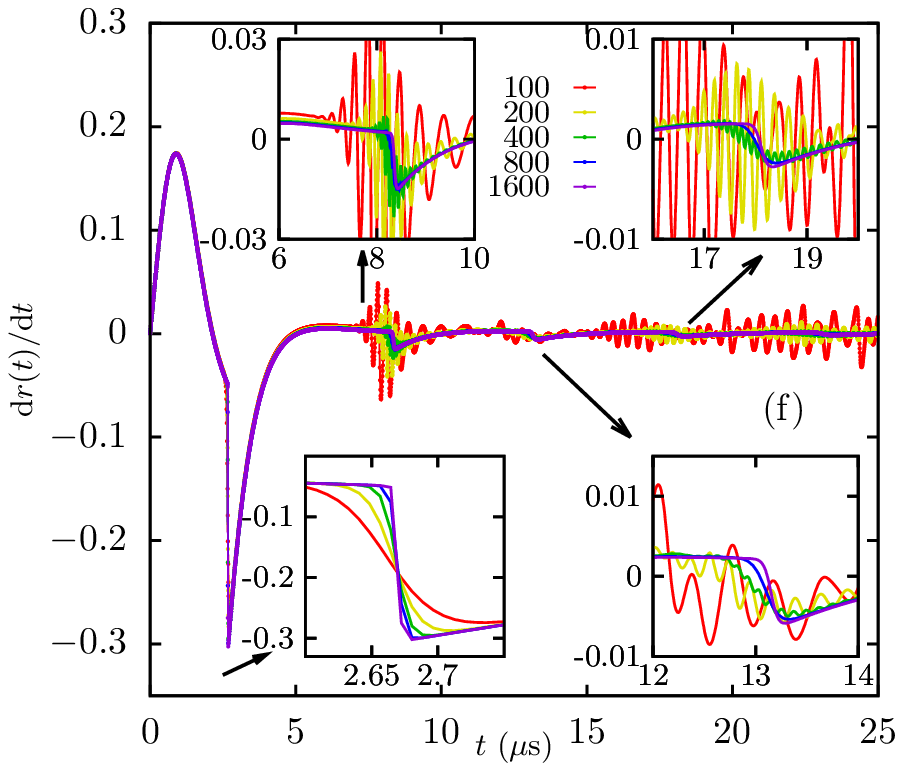}
\end{tabular}
\caption{(a-c) Rate function $r_N(t)$ of the PPRP $\mathcal{L}(t)$, Eq. \eqref{eq:erre}, for a quench from (a) $\lambda_i=2.535$, (b) $\lambda_{i}=4$ and (c) $\lambda_{i}=7.437$. All quenches end at $\lambda_{f}=1.6$. (d-f) Time derivatives $r'_N(t)$ corresponding to $r_{N}(t)$ in panels (a-c), respectively. The average energy of the quench is (a,d) $\epsilon=-1.07<\epsilon_{c}$, (b,e) $\epsilon=\epsilon_{c}=-1$ and (c,f) $\epsilon=-0.92>\epsilon_{c}$. The insets show magnifications of the finite-size scaling. Several values of the system size $j$ are indicated. The initial state considered for the quench has $\alpha=1/2$ and $\phi=0$. }
\label{fig:paneldptii}
\end{figure*}
\end{center}

\begin{center}
\begin{figure*}[h]
\hspace{-0.3cm}
\begin{tabular}{c c}
\includegraphics[width=0.45\textwidth]{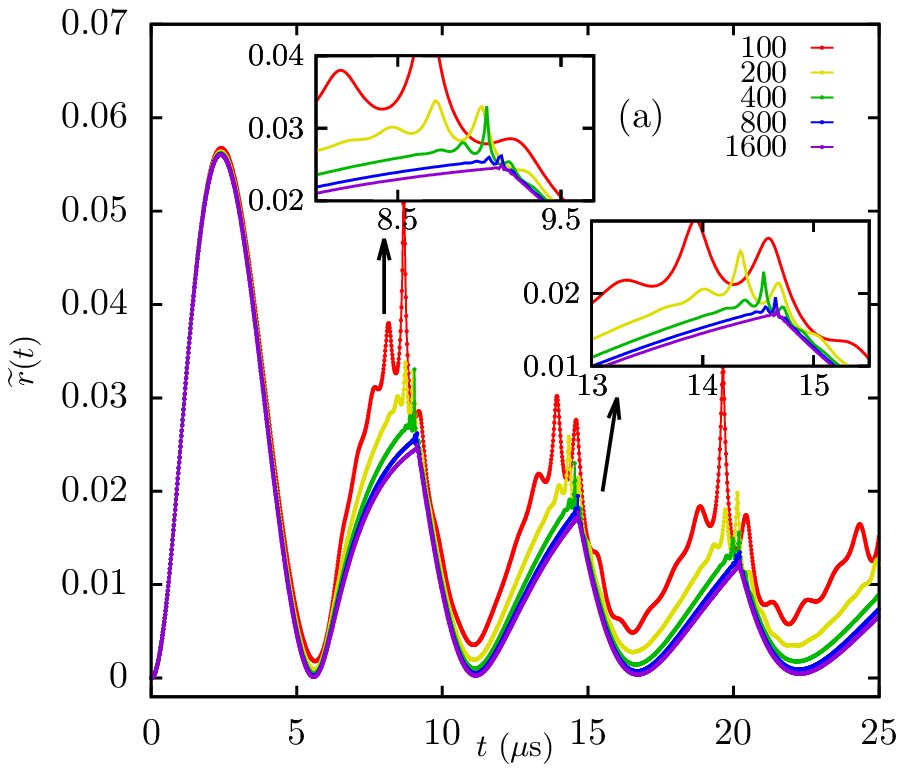} & \includegraphics[width=0.45\textwidth]{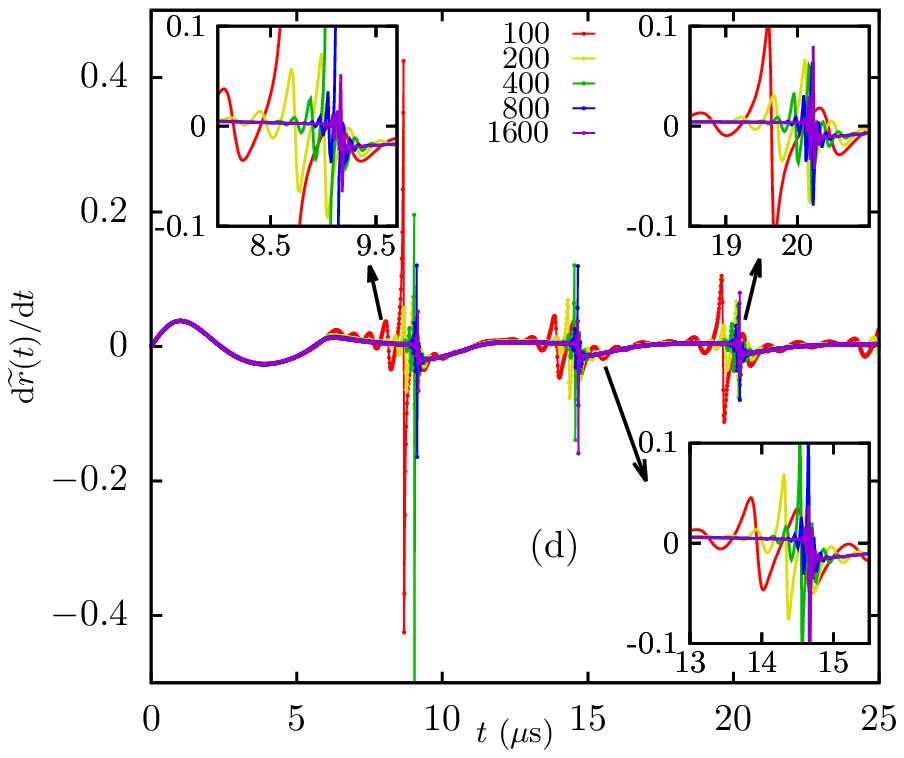} \\ \includegraphics[width=0.45\textwidth]{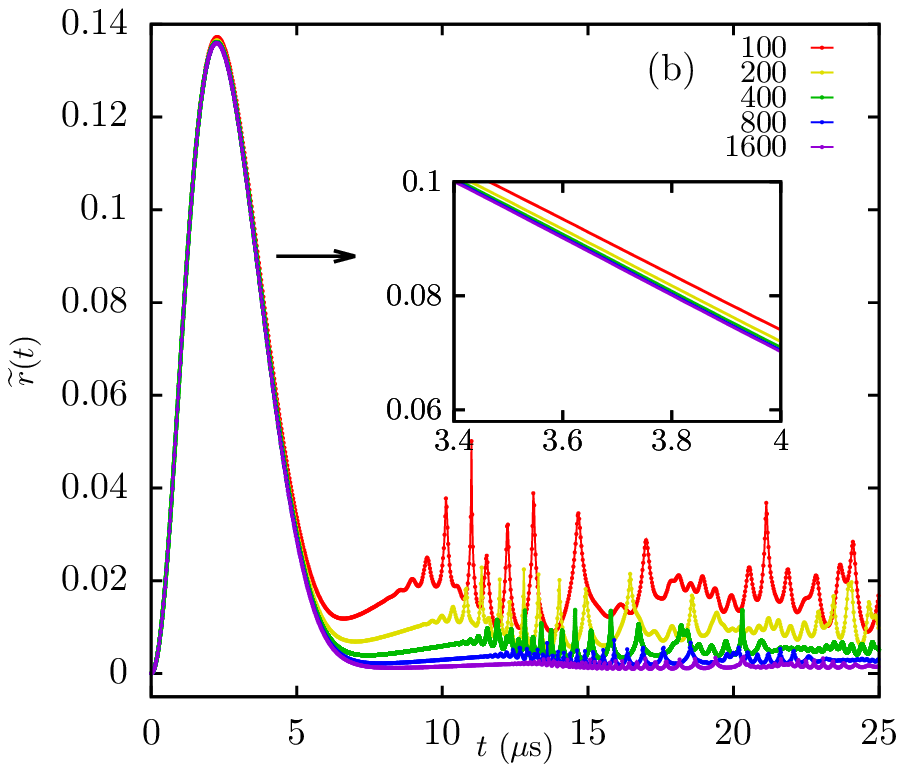} & \includegraphics[width=0.45\textwidth]{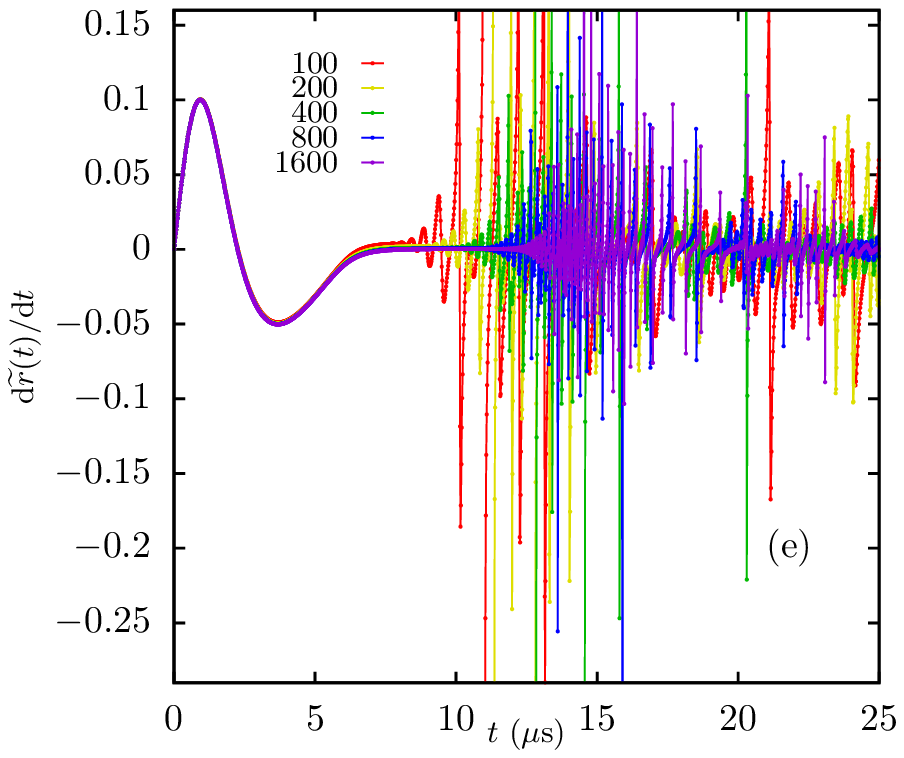} \\ \includegraphics[width=0.45\textwidth]{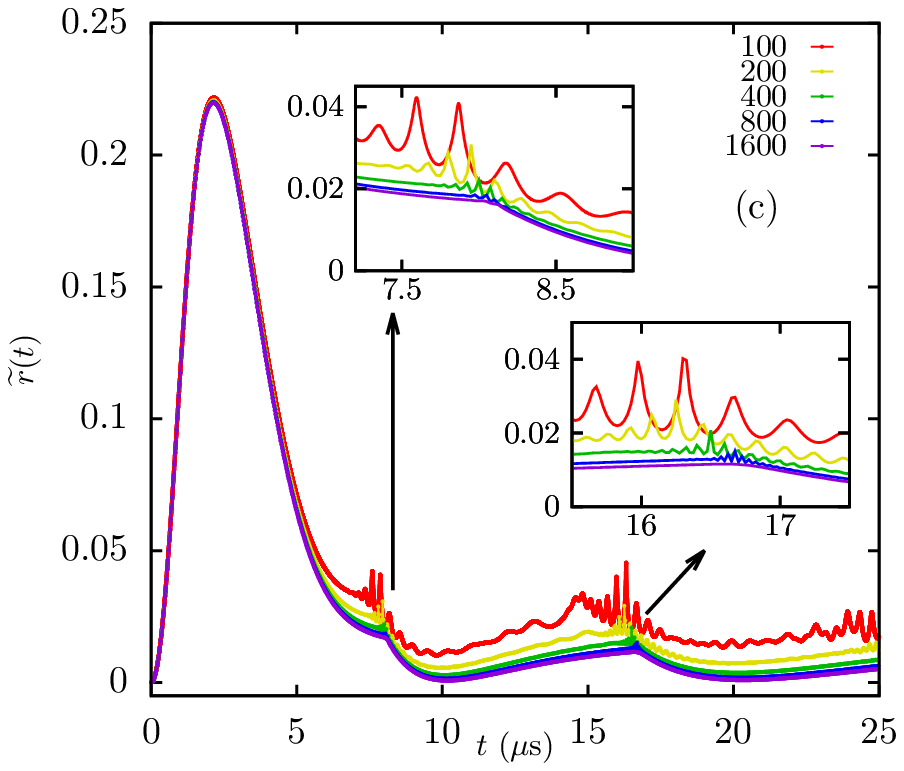} & \includegraphics[width=0.45\textwidth]{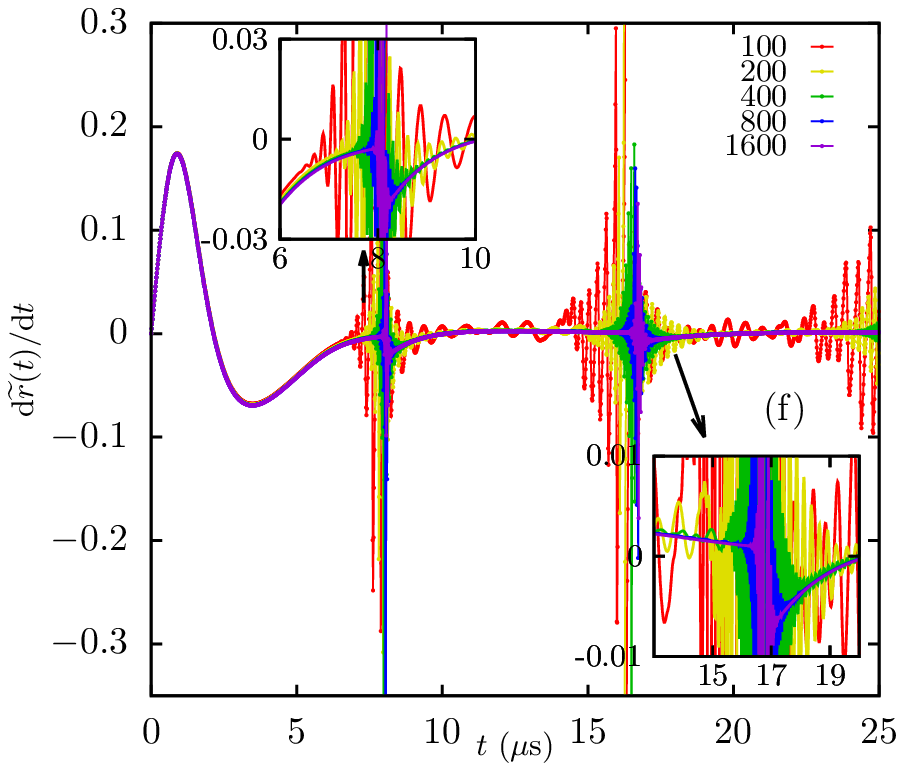}
\end{tabular}
\caption{Same results as in Fig. \ref{fig:paneldptii} but for the rate function of the survival probability, $\widetilde{r}(t)$, Eq. \eqref{eq:erretilde}.} 
\label{fig:panelsp}
\end{figure*}
\end{center}

In this section we provide a numerical analysis of the return probability, $\mathcal{L}(t)$, Eq. \eqref{eq:echo}, and the survival probability, $\textrm{SP}(t)$, Eq. \eqref{eq:survivalprobability}, after a quench. We start with an initial state of the form Eq. \eqref{eq:initialstate} with $\alpha=1/2$ and $\phi=0$, at a certain value of the control parameter $\lambda_{i}>\lambda_{c}$. The quench $\lambda_{i}\to\lambda_{f}$ always ends at $\lambda_{f}=1.6$, the final energy of the quench depending on $\lambda_{i}$ only. After the quench, the return probability $\mathcal{L}(t)$, Eq. \eqref{eq:echo}, is computed at different times, and then $r_{N}(t)$, Eq. \eqref{eq:erre}, is calculated. The non-analytic behavior of $r_{N}(t)$ at certain critical times, if they exist, is best captured by its first derivative $\textrm{d}r_{N}(t)/\textrm{d}t$, which we also consider. In our numerical calculations we have worked with up to 500 significant figures\footnote{It should be noted that in order to obtain high-resolution results, algorithms with standard numerical precision are insufficient because the $r_{N}(t)$ quickly drops below the precision limit; this means that obtaining reliable critical times from $r_{N}(t)$ may be challenging.}. Results for $r_{N}(t)$ are shown in Fig. \ref{fig:paneldptii}(a-c), while $\textrm{d}r_{N}(t)/\textrm{d}t$ is depicted in Fig. \ref{fig:paneldptii}(d-f). Different system sizes, $j=N/2$, are indicated. In the case of the survival probability, we consider, in analogy with Eq. \eqref{eq:erre}, the following rate function: 

\beq\label{eq:erretilde}
\widetilde{r}_{N}(t)=-\frac{1}{N}\ln \textrm{SP}(t).
\eeq
Note that this rate function is simply $\widetilde{r}_{N}(t)=2\textrm{Re}\,[g(t)]$. The corresponding results are shown in Fig. \ref{fig:panelsp}. In both Fig. \ref{fig:paneldptii} and Fig. \ref{fig:panelsp}, the first row shows the results for a quench with an average energy $\epsilon=-1.07<\epsilon_{c}$, which is below the ESQPT, i.e., the ESQPT has not been crossed. In the second row, the average energy coincides exactly with the ESQPT critical energy, $\epsilon_{c}=-1$. Finally, in the third row, the average energy is above the ESQPT, $\epsilon=-0.92>\epsilon_{c}$, so the ESQPT has been crossed. 

If $\epsilon<\epsilon_{c}$, we observe that results in Fig. \ref{fig:paneldptii}(a-b) and Fig. \ref{fig:panelsp}(a-b) are the same. As we have shown in the previous section, this is a consequence of the conservation of $\hatmath{C}$ in the TL. Some small differences between these figures can be observed only for small values of $j$, reflecting the fact that $\hatmath{C}$ is only strictly constant in the TL \cite{Corps2021}. We observe a set of oscillations in $r_{N}(t)$ and $\widetilde{r}_{N}(t)$, with some maxima. At first sight, the nature of some of these maxima is different. The first maximum at $t\approx 3$ does not show any relevant feature, as it appears to be smooth. However, the remaining maxima are apparently much sharper as $j$ increases, and have the appearance of kinks. Although this representation would hint to non-analytic behavior in $r_{N}(t)$ in the TL, and thus to a dynamical phase transition at certain critical times $t^{*}$, the main mechanism for this phenomenon is forbidden by our analytical results from the previous section since $\epsilon<\epsilon_{c}$. In fact, the derivatives shown in Figs. \ref{fig:paneldptii}(d) and \ref{fig:panelsp}(d) do not feature any clear scaling behavior around these kinks. These derivatives are magnified for clarity in the various insets of each panel, indicated by a black arrow. This is the so-called anomalous \cite{Homrighausen2017} phase of the DPT-II because all kinks in $r_{N}(t)$ and $\widetilde{r}_{N}(t)$ appear after the first minimum of these functions. Results in \cite{Homrighausen2017} suggest that the farther the final energy of the quench from $E_{c}$, the larger the number of smooth local maxima before the first kink appears. 

If $\epsilon>\epsilon_{c}$, the behavior of $r_{N}(t)$ and $\widetilde{r}_{N}(t)$ is different, as exemplified in Figs. \ref{fig:paneldptii}(c) and \ref{fig:panelsp}(c) and expected from our analytical results. In the case of $r_{N}(t)$ we observe some non-analytic points, whose derivatives, showed in Fig. \ref{fig:paneldptii}(f), display a scaling typical of a phase transition as the system size increases. These derivatives approach a clear discontinuous behavior as $j$ increases. This bears some resemblance to first order phase transitions because it is the first derivative of $r_{N}(t)$ and in the TL, $\lim_{N\to\infty}\textrm{d}r_{N}/\textrm{d}t$, which is discontinuous at certain critical times $t=t_{*}$. However, the singularities observed for $\widetilde{r}_{N}(t)$ in Fig. \ref{fig:panelsp}(c,f) are similar to those in the case $\epsilon<\epsilon_{c}$ [Fig. \ref{fig:panelsp}(a,d)] in that the derivatives $\textrm{d}\widetilde{r}_{N}(t)/\textrm{d}t$ do not show a clear scaling behavior when $N$ is increased. This is the regular phase of the DPT-II, because in both cases the kinks appear before the first minimum in $r_{N}(t)$ or $\widetilde{r}_{N}(t)$. 

Finally, if $\epsilon=\epsilon_{c}$, the behavior of $r_{N}(t)$ and $\widetilde{r}_{N}(t)$ is also different, as shown in Figs. \ref{fig:paneldptii}(b) and \ref{fig:panelsp}(b). The most relevant difference between $r_{N}(t)$ and $\widetilde{r}_{N}(t)$ in this critical case is that $\textrm{d}r_{N}/\textrm{d}t$ becomes discontinuous when $N\to\infty$ around $t_{*}\approx 3.75$ [Fig. \ref{fig:paneldptii}(e)], while such non-analytic point completely disappears in $\textrm{d}\widetilde{r}_{N}(t)/\textrm{d}t$ [Fig. \ref{fig:panelsp}(e)].

\begin{center}
\begin{figure}[h]
\hspace{-0.4cm}
\includegraphics[width=0.5\textwidth]{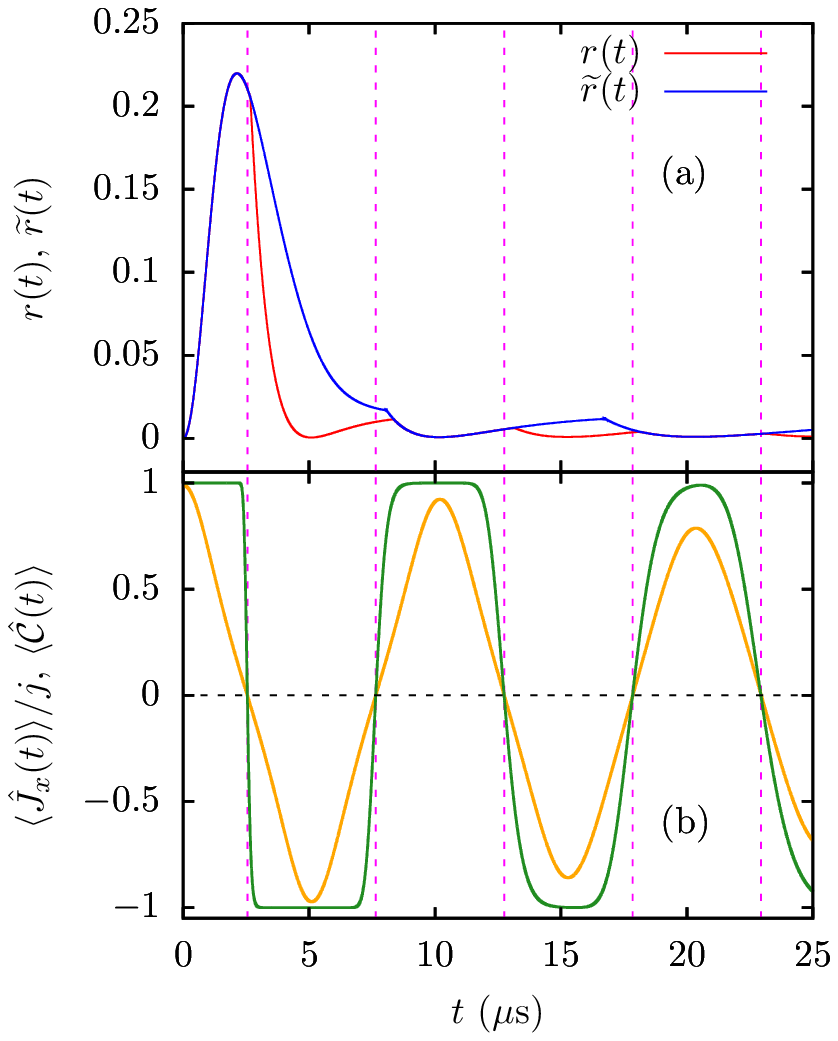} 
\caption{Comparison of indicators of DPT-I and DPT-II after a quench $\lambda_{i}=7.347\to\lambda_{f}=1.6$, which leads the initial state above the ESQPT critical energy, $\epsilon=-0.92>-1$. The initial state has $\alpha=1/2$ and $\phi=0$. Results correspond to $j=1600$. (a) $r_{N}(t)$ and $\widetilde{r}_{N}(t)$; (b) dynamics of $\hat{J}_{x}$ (orange) and $\hatmath{C}$ (green). In (b), the dashed horizontal line marks the value $0$. Dashed vertical lines signal the times when the expectation value of the order parameter vanishes, $\langle\hat{J}_{x}(t)\rangle=0$. }
\label{fig:comp}
\end{figure}
\end{center}

We also study the link between DPTs-II and the zeros of the order parameter of DPTs-I. In the case of the infinite-range transverse-field Ising model, it has been proposed that non-analytic times in the rate function and the time when the order parameters of DPT-I vanish are closely connected \cite{Halimeh2017,Homrighausen2017}. In Fig. \ref{fig:comp}(a) we show $r_{N}(t)$ and $\widetilde{r}_{N}(t)$ for $\lambda_{i}=7.437$ and $j=1600$, and in Fig. \ref{fig:comp}(b) we show $\langle \hat{J}_{x}(t)\rangle$ and $\langle \hatmath{C}(t)\rangle$ for the same quench. Therefore, these results correspond to the regular dynamical phase. We can see that the non-analytical points in $r_{N}(t)$ are \textit{close} to the times when the order parameters of DPT-I vanish, $\langle \hat{J}_{x}(t)\rangle=0$ and $\langle \hatmath{C}(t)\rangle=0$, but irregular deviations are also clear in the figure. Therefore, it seems that neither these results nor our theory are enough to conclude whether this apparent correlation is caused by a common mechanism or not. Anyhow, it is worth remarking that non-analytical points in $r_{N}(t)$ occur either when $r_{N}(t)$ and $\widetilde{r}_{N}(t)$ separate or when they become equal again. In terms of $f_{\pm}(t)$, defined in Eq. \eqref{eq:fplusminus}, this means that the first non-analytical point in $r_{N}(t)$ happens at the time $t$ when $f_{+}(t)$ and $f_{-}(t)$ separate (which cannot happen if $E<E_{c}$, as in this case $f_{+}(t)=f_{-}(t)$, $\forall t$). The second non-analytical point occurs when $f_{+}(t)$ and $f_{-}(t)$ coincide again, and so on. However, the kinks observed in $\widetilde{r}_{N}(t)$ are linked neither to the zeros of $\langle \hat{J}_{x}(t)\rangle$ and $\langle \hatmath{C}(t)\rangle$, nor to the behavior of $f_{\pm}(t)$. 

\subsection{Discussion and open questions}\label{sec:discussiondptii}

Our results allow us to formulate a classification of DPTs-II in collective quantum systems in terms of the region of the excited-state energy diagram where the quench ends: below, above, or at the ESQPT critical energy. 

(i) \textit{Anomalous phase,} $E<E_{c}$. In this phase, we have shown analytically that $\textrm{SP}(t)=\mathcal{L}(t)$ in the TL. Because $f_{+}(t)=f_{-}(t)$, the main mechanism for the appearance of kinks in $\mathcal{L}(t)$ is forbidden. Some kinks appear in $\textrm{SP}(t)$ and $\mathcal{L}(t)$ at certain times, but its behavior in the TL is unclear as they fail to show a finite-size scaling typical of quantum phase transitions. The first of these kinks appears after the first maximum in $r(t)$ and $\widetilde{r}(t)$, and the farther the final energy from $E_{c}$, the more smooth local maxima in these functions before the first kink appears. 

(ii) \textit{Regular phase,} $E>E_{c}$. In this phase, the survival probability $\textrm{SP}(t)$ and the echo $\mathcal{L}(t)$ do not coincide in general, $\textrm{SP}(t)\neq\mathcal{L}(t)$. Kinks appear in the rate function of $\mathcal{L}(t)$. The finite-size scaling of $\textrm{d}r(t)/\textrm{d}t$ is typical of quantum phase transitions and strongly suggests a true discontinuity in the TL. These kinks occur at the times when $\textrm{SP}(t)$ and $\mathcal{L}(t)$ separate or coincide again. Kinks are also present in $\textrm{SP}(t)$, but at different times than in $\mathcal{L}(t)$; unlike $\mathcal{L}(t)$, the scaling behavior of the rate function of $\textrm{SP}(t)$ seems unclear in the TL. 

(iii) \textit{Critical line,} $E=E_{c}$. Kinks appear in $\mathcal{L}(t)$ because the quench still populates some eigenstates with energy $E>E_{c}$. Therefore, our proof for $\textrm{SP}(t)=\mathcal{L}(t)$ when $E<E_{c}$ no longer applies and $\Omega_{+}(t)$ and $\Omega_{-}(t)$ may intersect at certain critical times. There are no kinks in $\textrm{SP}(t)$: one maximum in its rate function $\widetilde{r}(t)$ is observed at short times, and it then becomes highly noisy for long times. This suggests that the nature of the non-analyticities in $\textrm{SP}(t)$ may change precisely at the ESQPT.

\section{Conclusions}\label{sec:conclusions}
In this paper we provide a theory of DPTs in collective quantum systems, i.e., many-body quantum systems with infinite-range interaction. Our results are exemplified in the fully connected transverse-field Ising model; however, they remain valid also for an important class of collective many-body systems.  We have shown that two kinds of DPTs characterized by different phenomenology, DPTs-I and DPTs-II, are closely linked in these systems to ESQPTs. In the end, we have established the phase diagram in Fig. \ref{fig:phasediagram} for a large class of well-known collective systems exhibiting QPTs, ESQPTs, DPTs-I and DPTs-II. 

Regarding DPTs-I, we provide an understanding of the order parameters, $\overline{m}$, and describe their values across the critical point through a generalization of the standard microcanonical ensemble. This is the first main result of this paper. In particular, we show that the typical order parameters can only be non-zero for quantum quenches leading the initial state below the ESQPT criticality, i.e., it is possible that $\overline{m}\neq 0$ for $E<E_{c}$, whereas they always become zero (in the TL) when the quench leads the initial state above the ESQPT: $\overline{m}=0$ always if $E>E_{c}$. The phases demarcated by the ESQPT are characterized by markedly different thermodynamic properties. One of the phases, $E<E_{c}$, is characterized by the existence of three non-commuting charges, $\hat{\Pi}$, $\hatmath{C}$, and $\hatmath{K}$, allowing parity-broken long-time averages. By contrast, the parity symmetry is restored in the phase defined by $E>E_{c}$, where $\hat{\Pi}$ is the only remaining conserved charge. All of these features are accounted for by a generalization of the microcanonical ensemble. DPTs-I are thus associated with a mechanism of information erasure, whereby the details of the initial state are lost upon crossing the ESQPT and cannot be recovered by a backward quench protocol.  The non-equilibrium dynamics resulting from quantum quenches are also analyzed. As the system size increases, the quantum dynamics of the system follows the semiclassical expectations up to a time that grows with system size as a power-law if the quench ends below or above the ESQPT; yet, this scaling changes dramatically for quenches ending exactly at the ESQPT, turning into a logarithmic law instead. 

Regarding DPTs-II, we have shown analytically that the main mechanism for  non-analyticities in the rate function $r(t)$ of the parity-projected return probability $\mathcal{L}(t)$ can only happen within the phase with restored symmetry, $E>E_{c}$, while it is forbidden in the broken-symmetry phase, $E<E_{c}$. This is the second main result of this paper. We also show analytically that the usual definition of the survival probability $\textrm{SP}(t)$ coincides with the return probability $\mathcal{L}(t)$ only within the broken-symmetry phase. A numerical investigation suggests that if $E>E_{c}$, $\textrm{d}r(t)/\textrm{d}t$ becomes discontinuous in the TL at certain critical times, but the scaling of $\textrm{d}\widetilde{r}(t)/\textrm{d}t$ at any energy and that of $\textrm{d}r(t)/\textrm{d}t$ if $E<E_{c}$ is inconclusive. In the end, the regular and anomalous dynamical phases associated with DPTs-II can be formulated in terms of the energy of the quenched state starting from the broken-symmetry phase: when $E<E_{c}$, this gives rise to an anomalous \cite{Halimeh2017,Homrighausen2017} phase while when $E>E_{c}$ one finds a so-called regular phase. In the regular phase, $\textrm{d}r(t)/\textrm{d}t$ becomes discontinuous in the TL at certain times, while in the anomalous phase there are possible kinks in both $\textrm{d}r(t)/ \textrm{d}t$ and $\textrm{d}\widetilde{r}(t)/\textrm{d}t$. Thus, it is precisely the ESQPT at $E=E_{c}$ that triggers the change of phase type in these systems. 

A natural continuation of this work is to look for extensions of our systematic analysis for collective systems to quantum spin chains with finite-range interactions, where the concept of ESQPT has not been established due to the lack of a proper classical limit. Results in \cite{Halimeh2017,Lang2018} show that typical features of collective models, like the anomalous dynamical phase, also appear if the interaction between spins is long-range enough. This may open the door to ESQPT-esque behavior in such models. We hope to tackle this problem in the near future.

In closing, our work provides a unification of the concept of dynamical phase transition and that of excited-state quantum phase transitions in collective quantum systems, which should stimulate further research both on the theoretical and experimental ends.

\begin{acknowledgments}
We gratefully acknowledge discussions with P. P\'{e}rez-Fern\'{a}ndez and J. Dukelsky. This work has been supported by the Spanish grant PGC-2018-094180-B-I00 funded by Ministerio de Ciencia e Innovaci\'{o}n/Agencia Estatal de Investigaci\'{o}n MCIN/AEI/10.13039/501100011033 and FEDER "A Way of Making Europe". A. L. C. acknowledges financial support from `la Caixa' Foundation (ID 100010434) through the fellowship LCF/BQ/DR21/11880024.
\end{acknowledgments}


\begin{thebibliography}{100}

\bibitem{Gring2012} M. Gring, M. Kuhnert, T. Langen, T. Kitagawa, B. Rauer, M. Schreitl, I. Mazets, D. Adu Smith, E. Demler, and J. Schmiedmayer, \textit{Relaxation and Prethermalization in an Isolated Quantum System,} Science \textbf{337}, 1318 (2012).

\bibitem{Hofferberth2007} S. Hofferberth, I. Lesanovsky, B. Fischer, T. Schumm, and J. Schmiedmayer, \textit{Non-equilibrium coherence dynamics in a one-dimensional Bose gases,} Nature \textbf{449}, 324 (2007).

\bibitem{Hild2014} S. Hild, T. Fukuhara, P. Schau{\ss}, J. Zeiher, M. Knap, E. Demler, I. Bloch, and C. Gross, \textit{Far-from-equilibrium spin transport in Heisenberg quantum magnets,} Phys. Rev. Lett. \textbf{113}, 147205 (2014).

\bibitem{Yuznashyan2006} E. A. Yuzbashyan, O. Tsyplyatyev, and B. L. Altshuler, \textit{Relaxation and persistent oscillations of the order parameter in fermionic condensates,} Phys. Rev. Lett. \textbf{96}, 097005 (2006).

\bibitem{Baumann2010} K. Baumann, C. Guerlin, F. Brennecke, and T. Esslinger, \textit{Dicke quantum phase transition with a superfluid gas in an optical cavity}, Nature \textbf{464}, 1301 (2010).

\bibitem{Muniz2020} J. A Muniz, D. Barberena, R. J. Lewis-Swan, D. J. Young, J. R. K. Cline, A. M. Rey, and J. K. Thompson, \textit{Exploring dynamical phase transitions with cold atoms in an optical cavity}, Nature \textbf{580}, 602 (2020).

\bibitem{Chu2020} A. Chu, J. Will, J. Arlt, C. Klempt, and A. M. Rey, \textit{Simulation of $XXZ$ Spin Models Using Sideband Transitions in Trapped Bosonic Gases}, Phys. Rev. Lett. \textbf{125}, 240504 (2020).

\bibitem{Eckstein2008} M. Eckstein and M. Kollar, \textit{Nonthermal Steady States after an Interaction Quench in the Falicov-Kimball Model}, Phys. Rev. Lett. \textbf{100}, 120404 (2008).

\bibitem{Moeckel2008} M. Moeckel and S. Kehrein, \textit{Interaction Quench in the Hubbard Model}, Phys. Rev. Lett. \textbf{100}, 175702 (2008).

\bibitem{Eckstein2009} M. Eckstein, M. Kollar, and P. Werner, \textit{Thermalization after an Interaction Quench in the Hubbard Model}, Phys. Rev. Lett. \textbf{103}, 056403 (2009).

\bibitem{Sciolla2011} B. Sciolla and G. Biroli, \textit{Dynamical transitions and quantum quenches in mean-field models}, J. Stat. Mech. (2011) P11003.

\bibitem{Zhang2017} J. Zhang, G. Pagano, P. W. Hess, A. Kyprianidis, P. Becker, H. Kaplan, A. V. Gorshkov, Z.-X. Gong, and C. Monroe, \textit{Observation of a many-body dynamical phase transition with a 53-qubit quantum simulator}, Nature \textbf{551}, 601 (2017).

\bibitem{Smale2019} S. Smale, P. He, B. A. Olsen, K. G. Jackson, H. Sharum, S. Trotzky, J. Marino, A. M. Rey, and J. H. Thywissen, \textit{Observation of a Transition Between Dynamical Phases in a Quantum Degenerate Fermi Gas}, Science Advances \textbf{5}, eaax1568 (2019).

\bibitem{Tian2020} T. Tian, H.-X. Yang, L.-Y. Qiu, H.-Y. Liang, Y.-B. Yang, Y. Xu, and L.-M. Duan, \textit{Observation of Dynamical Quantum Phase Transitions with Correspondence in an
Excited State Phase Diagram}, Phys. Rev. Lett. \textbf{124}, 043001 (2020).

\bibitem{Halimeh2017prethermalization} J. C. Halimeh, V. Zauner-Stauber, I. P. McCulloch, I. de Vega, U. Schollw\"{o}ck, and M. Kastner, \textit{Prethermalization and persistent order in the absence of a thermal phase transition}, Phys. Rev. B \textbf{95}, 024302 (2017).

\bibitem{Sciolla2013} B. Sciolla and G. Biroli, \textit{Quantum quenches, dynamical transitions, and off-equilibrium quantum criticality}, Phys. Rev. B \textbf{88}, 201110(R) (2013).

\bibitem{Alvarez2006} G. A. \'{A}lvarez, E. P. Danieli, P. R. Levstein, and H. M. Pastawski, \textit{Environmentally induced quantum dynamical phase transition in the spin swapping operation}, J. Chem. Phys. \textbf{124}, 194507 (2006).

\bibitem{Mori2018} T. Mori, T. N. Ikeda, E. Kaminishi, and M. Ueda, \textit{Thermalization and prethermalization in isolated quantum systems: A theoretical overview}, J. Phys. B: At. Mol. Opt. Phys. \textbf{51}, 112001 (2018).

\bibitem{Prigogine1971} P. Glansdorff and I. Progogine, \textit{Thermodynamic Theory of Structure, Stability and Fluctuations}, Wiley, New York (1971).

\bibitem{Marino2022} J. Marino, M. Eckstein, M. S. Foster, and A. M. Rey, \textit{Dynamical phase transitions in the collisionless pre-thermal states of isolated quantum systems: theory and experiments}, arXiv:2201.09894 (2022).

\bibitem{Heyl2018} M. Heyl, \textit{Dynamical quantum phase transitions: a review},  	Rep. Prog. Phys. \textbf{81}, 054001 (2018)

\bibitem{Heyl2013} M. Heyl, A. Polkovnikov, and S. Kehrein, \textit{Dynamical Quantum Phase Transitions in the Transverse-Field Ising Model}, Phys. Rev. Lett. \textbf{110}, 135704 (2013).

\bibitem{Heyl2014} M. Heyl, \textit{Dynamical Quantum Phase Transitions in Systems with Broken-Symmetry Phases}, Phys. Rev. Lett. \textbf{113}, 205701 (2014).

\bibitem{Jurcevic2017} P. Jurcevic, H. Shen, P. Hauke, C. Maier, T. Brydges, C. Hempel, B. P. Lanyon, M. Heyl, R. Blatt, and C. F. Roos, \textit{Direct Observation of Dynamical Quantum Phase Transitions in an Interacting Many-Body System}, Phys. Rev. Lett. \textbf{119}, 080501 (2017).

\bibitem{Homrighausen2017} I. Homrighausen, N. O. Abeling, V. Zauner-Stauber, and J. C. Halimeh, \textit{Anomalous dynamical phase in quantum spin chains with long-range interactions}, Phys. Rev. B \textbf{96}, 104436 (2017).

\bibitem{Halimeh2017} J. C. Halimeh and V. Zauner-Stauber, \textit{Dynamical phase diagram of quantum spin chains with long-range interactions}, Phys. Rev. B \textbf{96}, 134427 (2017).

\bibitem{Heyl2019} M. Heyl, \textit{Dynamical quantum phase transitions: a survey}, EPL \textbf{125}, 26001 (2019).

\bibitem{Nicola2021} S. De Nicola, A. A. Michilidis, and M. Serbyn, \textit{Entanglement View of Dynamical Quantum Phase Transitions}, Phys. Rev. Lett. \textbf{126}, 040602 (2021).

\bibitem{Schmitt2015} M. Schmitt and S. Kehrein, \textit{Dynamical quantum phase transitions in the Kitaev honeycomb model}, Phys. Rev. B \textbf{92}, 075114 (2015).

\bibitem{Bhattacharya2017} U. Bhattacharya and A. Dutta, \textit{Emergent topology and dynamical quantum phase transitions in two-dimensional closed quantum systems}, Phys. Rev. B \textbf{96}, 014302 (2017).

\bibitem{Karrasch2013} C. Karrasch and D. Schuricht, \textit{Dynamical phase transitions after quenches in nonintegrable models}, Phys. Rev. B \textbf{87}, 195104 (2013).

\bibitem{Peng2015} X. Peng, H. Zhou, B.-B. Wei, J. Cui, J. Du, and R.-B. Liu, \textit{Experimental Observation of Lee-Yang Zeros}, Phys. Rev. Lett. \textbf{114}, 010601 (2015).

\bibitem{Jafari2022} R. Jafari, A. Akbari, U. Mishra, and H. Johannesson, \textit{Floquet dynamical quantum phase transitions under synchronized periodic driving}, Phys. Rev. B \textbf{105}, 094311 (2022).

\bibitem{Naji2022} J. Naji, M. Jafari, R. Jafari, and A. Akbari, \textit{Dissipative Floquet dynamical quantum phase transition}, Phys. Rev. A \textbf{105}, 022220 (2022).

\bibitem{Jafari2019} R. Jafari, \textit{Dynamical Quantum Phase Transition and Quasi Particle Excitation}, Scientific reports \textbf{9} (1), 2871 (2019).

\bibitem{Mishra2020} U. Mishra, R. Jafari, and A. Akbari, \textit{Disordered Kitaev chain with long-range pairing: Loschmidt echo revivals and dynamical phase transitions}, J. Phys. A: Math. Theor. \textbf{53} (2020) 375301.

\bibitem{Jafari2019prb} R. Jafari, H. Johannesson, A. Langari, and M. A. Martin-Delgado, \textit{Quench dynamics and zero-energy modes: The case of the Creutz model}, Phys. Rev. B \textbf{99}, 054302 (2019).

\bibitem{Jafari2021} R. Jafari and A. Akbari, \textit{Floquet dynamical phase transition and entanglement spectrum}, Phys. Rev. A \textbf{103}, 012204 (2021).

\bibitem{Vajna2014} S. Vajna and B. D\'{o}ra, \textit{Disentangling dynamical phase transitions from equilibrium phase transitions}, Phys. Rev. B \textbf{89}, 161105(R) (2014).

\bibitem{Andraschko2014} F. Andraschko and J. Sirker, \textit{Dynamical quantum phase transitions and the Loschmidt echo: A transfer matrix approach}, Phys. Rev. B \textbf{89}, 125120 (2014).

\bibitem{Lang2018concurrence} J. Lang, B. Frank and J. C. Halimeh, \textit{Concurrence of dynamical phase transitions at finite temperature in the fully connected transverse-field Ising model}, Phys. Rev. B \textbf{97}, 174401 (2018).

\bibitem{Zunkovic2018} B. Zunkovic, M. Heyl, M. Knap, and A. Silva, \textit{Dynamical Quantum Phase Transitions in Spin Chains with Long-Range Interactions: Merging Different Concepts of Nonequilibrium Criticality}, Phys. Rev. Lett. \textbf{120}, 130601 (2018).

\bibitem{Lang2018} J. Lang, B. Frank, and J. C. Halimeh, \textit{Dynamical Quantum Phase Transitions: A Geometric Picture}, Phys. Rev. Lett. \textbf{121}, 130603 (2018).

\bibitem{Puebla2020} R. Puebla, \textit{Finite-component dynamical quantum phase transitions}, Phys. Rev. B \textbf{102}, 220302(R) (2020).

\bibitem{Weidinger2017} S. A. Weidinger, M. Heyl, A. Silva and M. Knap, \textit{Dynamical quantum phase transitions in systems with continuous symmetry breaking}, Phys. Rev. B \textbf{96}, 134313 (2017).

\bibitem{Hashizume2022} T. Hashizume, I. P. McCulloch, and J. D. Halimeh, \textit{Dynamical phase transitions in the two-dimensional transverse-field Ising model}, Phys. Rev. Res. \textbf{4}, 013250 (2022).

\bibitem{Sehrawat2021} A. Sehrawat, C. Srivastava, and U. Sen, \textit{Dynamical phase transitions in the fully connected quantum Ising model: Time period and critical time}, Phys. Rev. B \textbf{104}, 085105 (2021).

\bibitem{Zunkovic2015} B. Žunkovic, A. Silva and M. Fabrizio, \textit{Dynamical phase transitions and Loschmidt echo in the infinite-range XY model}, Phil. Trans. R. Soc. A \textbf{374}: 20150160 (2015).

\bibitem{Lerose2019} A. Lerose, B. Zunkovic, J. Marino, A. Gambassi, and A. Silva, \textit{Impact of nonequilibrium fluctuations on prethermal dynamical phase transitions in long-range interacting spin chains}, Phys. Rev. B \textbf{99}, 045128 (2019).

\bibitem{Cejnar2021} P. Cejnar, , P. Str\'ansk\'y, M. Macek, and M. Kloc, \textit{Excited-state quantum phase transitions},  	J. Phys. A: Math. Theor. \textbf{54} (2021) 133001.

\bibitem{Corps2022letter} A. L. Corps and A. Rela\~{n}o, \textit{Theory of dynamical phase transitions in collective quantum systems}, Letter associated to this companion article, arXiv:2205.03443 [cond-mat.stat-mech].

\bibitem{Corps2021} A. L. Corps and A. Rela\~{n}o, \textit{Constant of Motion Identifying Excited-State Quantum Phases}, Phys. Rev. Lett. \textbf{127}, 130602 (2021).

\bibitem{Halimeh2020} J. C. Halimeh, M. V. Damme, V. Zauner-Stauber, and L. Vanderstraeten, \textit{Quasiparticle origin of dynamical quantum phase transitions}, Phys. Rev. Res. \textbf{2}, 033111 (2020).

\bibitem{Lipkin1965} H. Lipkin, N. Meshkov, and A. Glick, \textit{Validity of many-body approximation methods for a solvable model: (I). Exact solutions and perturbation theory}, Nucl. Phys. \textbf{62}, 188 (1965).

\bibitem{Dusuel2004} S. Dusuel and J. Vidal, \textit{Finite-Size Scaling Exponents of the Lipkin-Meshkov-Glick Model}, Phys. Rev. Lett. \textbf{93}, 237204 (2004).

\bibitem{Heiss2005} W. D. Heiss, F G Scholtz, and H. B. Geyer, \textit{The large N behaviour of the Lipkin model and exceptional points}, J. Phys. A: Math. Gen. \textbf{38}, 1843 (2005).

\bibitem{Leyvraz2005} F. Leyvraz and W. D. Heiss, \textit{Large-N Scaling Behavior of the Lipkin-Meshkov-Glick Model}, Phys. Rev. Lett. \textbf{95}, 050402 (2005).

\bibitem{Castanos2006} O. Casta\~{n}os, R. L\'opez-Pe\~{n}a, J. G. Hirsch, and E. L\'opez-Moreno, \textit{Classical and quantum phase transitions in the Lipkin-Meshkov-Glick model}, Phys. Rev. B \textbf{74}, 104118 (2006).

\bibitem{Ribeiro2007} P. Ribeiro, J. Vidal, and R. Mosseri, \textit{Thermodynamical Limit of the Lipkin-Meshkov-Glick Model}, Phys. Rev. Lett. \textbf{99}, 050402 (2007).

\bibitem{Ribeiro2008} P. Ribeiro, J. Vidal, and R. Mosseri, \textit{Exact spectrum of the Lipkin-Meshkov-Glick model in the thermodynamic limit
and finite-size corrections}, Phys. Rev. E \textbf{78}, 021106 (2008).

\bibitem{Relano2008} A. Rela\~{n}o, J. M. Arias, J. Dukelsky, J. E. Garc\'{\i}a-Ramos, and P. P\'erez-Fern\'andez, \textit{Decoherence as a signature of an excited-state quantum phase transition}, Phys. Rev. A \textbf{78}, 060102(R) (2008).

\bibitem{GarciaRamos2017} J. E. Garc\'{\i}a-Ramos, P. P\'erez-Fernández, and J. M. Arias, \textit{Excited-state quantum phase transitions in a two-fluid
Lipkin model}, Phys. Rev. C \textbf{95}, 054326 (2017).

\bibitem{Relano2016} A. Rela\~{n}o, M. A. Bastarrachea-Magnani, and S. Lerma-Hern\'{a}ndez, \textit{Approximated integrability of the Dicke model}, EPL \textbf{116}, 50005 (2016); M. A. Bastarrachea-Magnani, A. Rela\~{n}o, S. Lerma-Hern\'{a}ndez, B. L\'{o}pez-del-Carpio, J. Ch\'{a}vez-Carlos, and J. G. Hirsch, \textit{Adiabatic invariants for the regular region of the Dicke model}, J. Phys. A: Math. Theor. \textbf{50} 144002 (2017).

\bibitem{Puebla2016} R. Puebla, M.-J. Hwang, and M. B. Plenio, {\em Excited-state quantum phase transition in the Rabi model}, Phys. Rev. A {\bf 94}, 023835 (2016).

\bibitem{Hwang2015} M.-J. Hwang, R. Puebla, and M. B. Plenio, {\em Quantum Phase Transition and Universal Dynamics in the Rabi Model}, Phys. Rev. Lett. {\bf 115}, 180404 (2015).

\bibitem{Lobez2016} C. M. L\'{o}bez and A. Rela\~{n}o, \textit{Entropy, chaos, and excited-state quantum phase transitions in the Dicke model}, Phys. Rev. E \textbf{94}, 012140 (2016).

\bibitem{Bastarrachea2014} M. A. Bastarrachea-Magnani, S. Lerma-Hern\'andez, J. G. Hirsch, {\em Comparative quantum and semi-classical analysis of Atom-Field Systems I: density of states and excited-state quantum phase transitions}, Phys. Rev. A {\bf 89}, 032101 (2014).

\bibitem{Perez2011b} P. P\'erez-Fern\'andez,  A. Rela\~no, J. M. Arias, P. Cejnar, J. Dukelsky, and J. E. Garc\'{\i}a-Ramos, {\em Excited-state phase transition and onset of chaos in quantum optical models}, Phys. Rev. E {\bf 83}, 046208 (2011).

\bibitem{Perez2011} P. P\'erez-Fern\'andez, P. Cejnar, J. M. Arias, J. Dukelsky, J. E. Garc\'{\i}a-Ramos, and A. Rela\~no, {\em Quantum quench influenced by an excited-state phase transition}, Phys. Rev. A {\bf 83}, 033802 (2011).

\bibitem{Brandes2013} T. Brandes, \textit{Excited-state quantum phase transitions in Dicke superradiance models}, Phys. Rev. E \textbf{88}, 032133 (2013).

\bibitem{LewisSwan2021} R. J. Lewis-Swan, S. R. Muleady, D. Barberena, J. J. Bollinger, and A. M. Rey, \textit{Characterizing the dynamical phase diagram of the Dicke model via classical and quantum probes}, Phys. Rev. Res. \textbf{3}, L022020 (2021).

\bibitem{Kloc2017} M. Kloc, P. Str\'{a}nsk\'{y} and P. Cejnar, \textit{Quantum phases and entanglement properties of an extended Dicke model}, Ann. Phys. \textbf{382}, 85 (2017).

\bibitem{Kloc2018} M. Kloc, P. Str\'{a}nsk\'{y} and P. Cejnar, \textit{Quantum quench dynamics in Dicke superradiance models}, Phys. Rev. A \textbf{98}, 013836 (2018).

\bibitem{Wang2021} Q. Wang and F. P\'erez-Bernal, \textit{Signatures of excited-state
quamtum phase transitions in quantum many-body systems:
Phase space analysis}, Phys. Rev. E \textbf{104}, 034119 (2021).

\bibitem{Feldmann2021} P. Feldmann, C. Klempt, A. Smerzi, L. Santos, and M.
Gessner, \textit{Interferometric Order Parameter for Excited-State
Quantum Phase Transitions in Bose-Einstein Condensates},
Phys. Rev. Lett. \textbf{126}, 230602 (2021).

\bibitem{Relano2014} A. Rela\~{n}o, J. Dukelsky, P. P\'{e}rez-Fern\'{a}ndez, and J. M. Arias, \textit{Quantum phase transitions of atom-molecule Bose mixtures in a double-well potential}, Phys. Rev. E \textbf{90}, 042139 (2014).

\bibitem{Sachdev1999} S. Sachdev, {\em Quantum Phase Transitions}, Cambridge University Press (1999).

\bibitem{Puebla2013} R. Puebla, A. Rela\~{n}o, and J. Retamosa, \textit{Excited-state phase transition leading to symmetry-breaking steady states in the Dicke model}, Phys. Rev. A \textbf{87}, 023819 (2013).

\bibitem{Puebla2013b} R. Puebla, and A. Rela\~{n}o, \textit{Non-thermal excited-state quantum phase transitions}, EPL \textbf{104}, 50007 (2013).

\bibitem{Puebla2015} R. Puebla, and A. Rela\~{n}o, \textit{Irreversible processes without energy dissipation in an isolated Lipkin-Meshkov-Glick model}, Phys. Rev. E \textbf{92}, 012101 (2015).

\bibitem{Stransky2014} P. Str\'{a}nsk\'{y}, M. Macek, and P. Cejnar, \textit{Excited-state quantum phase transitions in systems with two degrees of freedom: Level density, level dynamics, thermal properties}, Ann. Phys. (N.Y.) \textbf{345}, 73 (2014).

\bibitem{Botet1982} R. Botet, R. Jullien, and P. Pfeuty, \textit{Size scaling for infinitely coordinated systems}, Phys. Rev. Lett. \textbf{49}, 478 (1982).

\bibitem{Caprio2008} M. A. Caprio, P. Cejnar and F. Iachello, \textit{Excited state quantum phase transitions in many-body systems}, Ann. Phys. (N.Y.) \textbf{323}, 1106 (2008).

\bibitem{Gutzwillerbook} M. C. Gutzwiller, Chaos in Classical and Quantum Mechanics (1th edition, Springer New York, NY).

\bibitem{Corps2022} A. L. Corps, R. A. Molina and A. Rela\~{n}o, \textit{Chaos in a deformed Dicke model}, J. Phys. A: Math. Theor. \textbf{55} (2022) 084001.

\bibitem{Corps2022arxiv} A. L. Corps and A. Rela\~{n}o, \textit{Energy cat states induced by a parity-breaking excited-state quantum phase transition}, Phys. Rev. A \textbf{105}, 052204 (2022).

\bibitem{Alessio2016} L. D'Alessio, Y. Kafri, A. Polkovnikov, and M. Rigol, \textit{From Quantum Chaos and Eigenstate Thermalization to Statistical Mechanics and Thermodynamics}, Adv. Phys. \textbf{65}, 239 (2016).

\bibitem{Santos2016} L. F. Santos, M. T\'{a}vora, and F. P\'{e}rez-Bernal, \textit{Excited-state quantum phase transitions in many-body systems with infinite-range interaction: Localization, dynamics, and bifurcation}, Phys. Rev. A \textbf{94}, 012113 (2016).

\bibitem{Santos2015} L. F. Santos and F. P\'{e}rez-Bernal, \textit{Structure of eigenstates and quench dynamics at an excited-state quantum phase transition}, Phys. Rev. A \textbf{92}, 050101(R) (2015).

\bibitem{Milburn1997} G. J. Milburn and J. Corney, \textit{Quantum dynamics of an atomic Bose-Einstein condensate in a double-well potential}, Phys. Rev. A \textbf{55}, 4318 (1997).

\bibitem{Reimann2008} P. Reimann, \textit{Foundations of statistical mechanics under experimentally realistic conditions}, Phys. Rev. Lett. \textbf{101}, 190403 (2008).

\bibitem{Srednicki1999} M. Srednicki, \textit{The approach to thermal equilibrium in quantized chaotic systems}, J. Phys. A: Math. Theor. \textbf{32}, 1163 (1999).

\bibitem{Guryanova2016} Y. Guryanova, S. Popescu, A. J. Short, R. Silva, and P. Skrzypczyk, \textit{Thermodynamics of quantum systems with multiple conserved quantities}, Nat. Comm. \textbf{7}, 12049 (2016).

\bibitem{Halpern2016} N. Y. Halpern, P. Faist, J. Oppenheim, and A. Winter, \textit{Microcanonical and resource-theoretic derivations of the thermal state of a quantum system with noncommuting charges}, Nat. Comm. \textbf{7}, 12051 (2016).

\bibitem{Halpern2020} N. Y. Halpern, M. E. Beverland, and A. Kalev, \textit{Noncommuting conserved charges in quantum many-body thermalization}, Phys. Rev. E \textbf{101}, 042117 (2020).

\bibitem{Jaynes1957} E. T. Jaynes, \textit{Information theory and statistical mechanics}, Phys. Rev. \textbf{106}, 620 (1957); \textit{Information theory and statistical mechanics II}, ibid \textbf{108}, 171 (1957). 

\bibitem{Rigol2007} M. Rigol, M. Dunjko, V. Yurovsky, and M. Olshanii, \textit{Relaxation in a Completely Integrable Many-Body Quantum System: an Ab Initio Study of the Dynamics of the Highly Excited States of 1D Hard-Core Bosons}, Phys. Rev. Lett. \textbf{98}, 050405 (2007). 

\bibitem{Vidmar2016} L. Vidmar and M. Rigol, \textit{Generalized Gibbs ensemble in integrable lattice models}, J. Stat. Mech. (2016) 064007. 

\bibitem{Touchette2009} H. Touchette, \textit{The large deviation approach to statistical mechanics}, Phys. Rep. \textbf{478}, 1 (2009).


\end{thebibliography}
\end{document}